\documentclass[a4paper,fleqn]{cas-sc}
\usepackage[authoryear]{natbib}
\usepackage{graphicx} 
\usepackage{adjustbox}
\usepackage{tabularx,makecell, multirow,booktabs,caption}
\usepackage{threeparttable}
\usepackage{subcaption}
\usepackage{changepage}   
\usepackage{booktabs}
\usepackage{pifont}
\usepackage[nice]{units}
\makeatletter
\newcommand*{\rom}[1]{\expandafter\@slowromancap\romannumeral #1@}
\makeatother

\newenvironment{myenv}{\begin{adjustwidth}{0.5cm}{}}{\end{adjustwidth}}
\usepackage{booktabs, array} 
\begin{document}
	\let\WriteBookmarks\relax
	\def\floatpagepagefraction{1}
	\def\textpagefraction{.001}
	
	\shorttitle{S. K. Chada et al./Transportation Research Part C}
	
	\shortauthors{S. K. Chada et~al.}
	
	\title [mode = title]{Evaluation of the Driving Performance and User Acceptance of a Predictive Eco-Driving Assistance System for Electric Vehicles}                      
	
	%
	\author[1]{Sai Krishna Chada}[type=editor,
	auid=000,
	orcid=0000-0003-2663-4645]
	
	\cormark[1]
	
	\ead{chada@eit.uni-kl.de}
	
	\credit{Conceptualization of this study, Methodology, Software}
	
	\affiliation[1]{organization={Institute of Electromobility, University of Kaiserslautern},
		city={67663 Kaiserslautern},
		country={Germany}}
	
	
	\author[1]{Daniel Görges}[]
	
	\credit{Data curation, Writing - Original draft preparation}
	
	\author%
	[2] {Achim Ebert}
	
	\affiliation[2]{organization={Human Computer Interaction Group, University of Kaiserslautern},
		city={67663 Kaiserslautern},
		country={Germany}}
	
	\author%
	[3] {Roman Teutsch}
	
	\affiliation[3]{organization={Institute for Mechanical and Automotive Design, University of Kaiserslautern},
		city={67663 Kaiserslautern},
		country={Germany}}
	
	\author[1]{Shreevatsa Puttige Subramanya}[]
	
	\cortext[cor1]{Corresponding author}
	
	\begin{abstract}
		In this work, a predictive eco-driving assistance system (pEDAS) with the goal to assist drivers in improving their driving style and thereby reducing the energy consumption in battery electric vehicles (BEVs) while enhancing the driving safety and comfort is introduced and evaluated. pEDAS in this work is equipped with two model predictive controllers (MPCs), namely reference-tracking MPC and car-following MPC, that use the information from onboard sensors, signal phase and timing (SPaT) messages from traffic light infrastructure, and geographical information of the driving route to compute an energy-optimal driving speed. An optimal speed suggestion and informative advice are indicated to the driver using a visual feedback. Moreover, the warning alerts during unsafe car-following situations are provided through an auditory feedback. pEDAS provides continuous feedback and encourages the drivers to perform energy-efficient car-following while tracking a preceding vehicle, travel at safe speeds at turns and curved roads, drive at energy-optimal speed determined using dynamic programming in freeway scenarios, and travel with a green-wave optimal speed to cross the signalized intersections at a green phase whenever possible. Furthermore, to evaluate the efficacy of the proposed eco-driving assistance system, user studies were conducted with 41 participants on a dynamic driving simulator. The objective analysis revealed that the drivers achieved mean energy savings up to 10\%, reduced the speed limit violations, and avoided unnecessary stops at signalized intersections by using pEDAS. Finally, the user acceptance of the proposed pEDAS was evaluated using the Technology Acceptance Model (TAM) and Theory of Planned Behavior (TPB). The results showed an overall positive attitude of users and that the perceived usefulness and perceived behavioral control were found to be the significant factors in influencing the behavioral intention to use pEDAS.
	\end{abstract}
	
	
	
	\begin{keywords}
		Model predictive control \sep Dynamic programming \sep Eco-driving \sep Battery electric vehicles \sep Technology acceptance model \sep Theory of planned behavior 
	\end{keywords}
	\maketitle
	
	\section{Introduction}
	According to \citet{GAM21}, the global automotive market is expected to accelerate at a compound annual growth rate of 3.71\% from 85.32 million in 2020 to 122.83 million units by 2030. With the rapid growth in the number of road vehicles worldwide, the transportation sector is confronted with the arising challenges of greenhouse gas (GHG) emissions and traffic congestion \citep{Barkenbus2010,Sullman2015}. 
	The legislative bodies worldwide have committed to address the alarming climate change challenge and set stringent regulatory standards to reduce the global CO2 emissions from road vehicles \citep{Barkenbus2010}. Furthermore, the transportation industry is transitioning towards sustainable future mobility solutions, thus resulting in a paradigm shift towards electrified powertrains. Battery electric vehicles (BEVs) are regarded as an environmentally friendly transportation solution to curb the global CO2 emissions with their contribution towards zero tailpipe emissions \citep{Neubauer2014}. However, the BEVs must be operated as efficiently as possible since the electric power produced from fossil sources such as coal and natural gas power plants is partly being used to charge batteries in the transition phase \citep{SandyThomas2012}.
	Moreover, the limited battery capacity and driving range in BEVs is a major challenge for their adoption among users \citep{Neumann2015}. 	
	To address these challenges and satisfy the arising customer needs, the energy efficiency of the BEVs must be further improved. This has strengthened the motivation to develop advanced technological solutions that could enhance the energy efficiency in vehicles. On the one hand, the global research is focused on the vehicle side, specifically on the integration of lightweight materials \citep{GonzalezPalencia2012}, improving the powertrain efficiency \citep{Shao2020}, and developing advanced battery technologies \citep{Neubauer2014}. On the other hand, road vehicles have seen an inclusion of several driver assistance systems \citep{Bengler2014} over the past decades. The driver assistance systems available in the market can be predominantly categorized into two segments. Firstly, the safety-oriented driver assistance systems that primarily focus on aiding the drivers in enhancing their driving safety and comfort. A few examples are reverse parking assist, blind spot assist, lane keeping assist and emergency braking assist \citep{Bengler2014}. Secondly, the energy-oriented driver assistance systems which mainly assist the drivers in reducing their fuel or energy consumption by either providing indications in a simplified manner on the vehicle instrument cluster, for instance the instantaneous fuel consumption during the trip, average fuel consumption after the trip and indication of optimal gear shift or
	by taking over the driving responsibility partially, for example, the predictive cruise control \citep{Park2013} that can anticipate the elevation changes in the driving route and activate a coasting strategy to improve the fuel-efficiency. Furthermore, automated energy-optimal longitudinal control in vehicles is studied previously in several works \citep{Alamdari2019}, including ours \citep{Chada2020,Chada2021_2}. To enhance the energy efficiency in vehicles in a cost-effective manner, an area of research that has received significant attention over the years in both academia and automotive industry, is eco-driving \citep{Barkenbus2010}. The term eco-driving is referred to as an interim policy to improve the driver’s driving style in an energy-efficient manner. Many studies in the past have reviewed the benefits and limitations of the eco-driving policies \citep{Barkenbus2010, Alam2014}. 
	
	In the past studies, several eco-driving strategies for drivers using the conventional internal combustion engine (ICE) vehicles were presented \citep{Barkenbus2010,Ayyildiz2017}. For instance, encouraging the drivers to avoid unnecessary acceleration and deceleration, upshift to avoid engine speeds above 2500 rpm in vehicles with manual transmission, reduce idling, travel with constant driving speeds, drive below the regulatory speed limits, anticipate traffic objects and traffic light (TL) signals are some of the existing eco-driving techniques \citep{Barkenbus2010}. Although a majority of the aforementioned techniques apply for BEVs as well, the drivers must adjust a few techniques they know from conventional ICE vehicles to BEVs. For example, BEVs do not require gear shifting and therefore the upshifting rule is not relevant anymore. Moreover, a study from \citet{Neumann2015} reported that the users after experiencing BEVs for three months demonstrated eco-driving strategies such as reduced use of auxiliary functions (air conditioning),  smooth usage of the accelerator pedal and learnt to moderately apply regenerative braking to improve the energy savings. 
	
	The BEV energy consumption in the real-world may be affected by several factors, for instance, travelling at higher road speeds might result in additional energy consumptions and can adversely affect the driving range in BEVs \citep{Wager2016}. Moreover, repeated interactions with a preceding vehicle can negatively influence the acceleration and deceleration behaviors in BEVs, that can in turn affect its energy consumption. In addition to the dense traffic in urban environments, another important element that can influence the BEV energy consumption is the traffic light (TL) signals. \citet{Rakha2003} claimed that the unnecessary stops at TL signals can lead to an increase in the energy consumption in vehicles up to approximately 10\%. Existing studies on eco-driving support systems use comparatively simple optimization methods and provide momentary recommendations to the drivers in less complex driving situations. There is a considerable need for research and evaluation in developing advanced eco-driving assistance systems that make predictive suggestions to drivers in a continuous fashion by accessing the surrounding information from on-board sensors, vehicle-to-vehicle (V2V) and vehicle-to-infrastructure (V2I) communications. Moreover, to successfully integrate new technologies into the transportation industry, the user acceptance plays a crucial role. The user acceptance is defined as the degree to which the user possesses a positive intention to use the technology in his or her daily life if it is made available \citep{Adell2009}. Therefore, exposing new eco-driving systems to the users during the early development phase can help to obtain an early feedback regarding the user acceptance and the necessary improvements as well.

To address the aforementioned research gaps, this work focuses on developing a predictive eco-driving assistance system (pEDAS) that assists the drivers to improve their driving behavior in an energy-efficient way by providing a continuous speed advisory feedback. The goal here is to provide recommendations to the driver while interacting with surrounding traffic objects, traffic light signals, traffic signs and curved roads. Moreover, the study aims at communicating the optimal advisory speed along with some additional information that can help the driver to easily interpret the speed recommendation through a human-machine interface (HMI). Another focus of this study is to evaluate the driving performance of the proposed pEDAS by conducting user studies on a dynamic driving simulator, in which the participants test the functionality of pEDAS on both urban and highway driving environments. Apparently, performing driver-in-the-loop tests using a driving simulator is crucial in the early development phase as this helps to drastically reduce the real-world testing times and is much safer as well \citep{Chada2021_1}. The ultimate goal of the study is to determine the key factors that influence the user acceptance of pEDAS through the subjective data collected from a survey questionnaire designed using state-of-the-art technology acceptance models. To the best knowledge of the authors, evaluation of a continuous predictive EDAS for battery electric vehicles in both an objective and subjective manner has not yet been addressed in the literature. 	

\section{Literature Review}
\label{sec:state_of_the_art}		
	\subsection{Related Work on Eco-driving}
	\label{sec:literature_EDAS}	
To encourage drivers to learn an energy-efficient driving behavior, several eco-driving programs were investigated in the past. A curated list of literature is presented in Table~\ref{tbl1}. Coaching the drivers on eco-driving techniques can contribute to reductions in fuel consumption of 5.5-16.9\% \citep{Beusen2009, Sullman2015, Beloufa2017, Ayyildiz2017, Ruben2021}. Many studies have pointed out that the drivers tend to forget the eco-driving techniques learned through training sessions and get back to their old driving habits in a short period, thus showcasing no long term benefits of such trainings \citep{Beusen2009, Beloufa2017}. In a driving simulator experiment by \citet{Daun2013}, displaying a few eco-driving advisory messages to the drivers has improved their fuel savings up to 12\%. However, in a consecutive experiment when the drivers were not provided with any advisory messages, it did not yield in any fuel saving benefits. This points out the necessity for a continuous eco-driving feedback system \citep{Daun2013,Vagg2013,Staubach2014,Zhao2015,Fors2015,Jamson2015,Gonzalez2021} to avail the long-term benefits in terms of fuel or energy savings. Using in-vehicle feedback systems that provide continuous suggestions in a visual, haptic and auditory manner, drivers were able to perform additional eco-driving behaviors such as coasting and reduced harsh acceleration \citep{McIlroy2017}. One of the negative impacts of an eco-driving feedback system is that it enforces drivers to focus on the suggestions and may increase the risk of accidents by causing driver distractions \citep{Alam2014}. It is therefore crucial to provide simple and clear indications that are easy for the drivers to interpret. An effective in-vehicle human-machine interface (HMI) plays a major role in improving the driver engagement with an eco-driving assistance system (EDAS).

The concept of predictive EDAS based on the topology (such as elevation, curvature and speed limits) of the route has been studied by \citep{Daun2013}. In this work, the efficacy of a predictive EDAS that provides the driver with situational and contextual advice to improve the fuel efficiency in heavy commercial vehicle is analysed using a driving simulator experiment. The study, however, relied on very few parameters for determining an optimal driving strategy. A recent work \citet{Alam2018} suggests that the prediction information of the surrounding traffic elements obtained from the wireless communication systems such as vehicle-to-vehicle (V2V) and vehicle-to-infrastructure (V2I) in combination with eco-driving techniques can help to reduce the CO2 emissions and traffic congestion in a network level. A few studies have demonstrated the benefits of using the signal phase and timing (SPaT) information from the TL signals to optimally plan the host vehicle trajectories at signalized intersections \citep{Asadi2011,Staubach2014}. In \citep{Staubach2014}, the authors developed an eco-driving support system to suggest an optimal speed with which the drivers can cross the immediate TL signals at green. Using the support system, drivers achieved average fuel savings up to 15.9-18.4\%. The limitation in their study is that the interactions with surrounding traffic were not considered. According to our recent findings \citep{Chada2020}, the energy savings can increase up to 26\% if the host vehicle tracks a green-wave optimal speed (GWOS). It is to be noted that these energy savings are realistic only in freeway scenarios where the influence due to other traffic objects is not present.   


In addition to minimizing the host vehicle's energy or fuel consumption, recent studies point towards EDAS that aim at improving driving safety and comfort. To address these multiple objectives, an optimal control problem \citep{Fleming2021} is solved using methods such as Pontryagin’s minimum principle \citep{Ozatay2017} and dynamic programming (DP) \citep{Lin2014, Maamria2016, Ozatay2017}. The authors have commonly pointed out that although these methods have shown to yield a global optimum solution, they are found to be computationally intensive, thus making them infeasible for the online implementation. Alternatively, it is possible to calculate the DP optimization problem offline at the beginning of the trip and the results of which can be used as a reference for online implementation. Such an approach is presented in \citet{Lin2014}, in which a DP algorithm uses the route topology information such as elevation, speed limits to calculate an optimal reference velocity for the entire driving route. To address the limitation of the aforementioned methods, a widely popular approach known as model predictive control (MPC) is used to facilitate online-control and effectively deal with state and input constraints. Several literature exist that implemented MPC to solve multi-objective optimization problems for advanced adaptive cruise control (ACC) concepts in automated driving \citep{Li2009, Weissmann2018, Jia2020}, however, very few studies extended the MPC framework to EDAS. For instance, \citet{Chen2022} proposed an EDAS that used a simplified model for the speed advisor to compute driving mode suggestions such as cruising, eco-roll, coasting and engine braking. Although their study presented some preliminary work on driver delay compensation using MPC, a major limitation is that the study was restricted to deceleration scenarios alone. The primary objective in our study is to develop a predictive eco-driving assistance system (pEDAS) using a two-level model predictive control framework with a goal to assist BEV drivers in the presence of surrounding traffic, traffic light signals, traffic signs and curved roads by providing a continuous feedback to improve their driving behavior in an energy-efficient manner.
		\begin{table}[width=1\linewidth,pos=tp]
		\caption{Summary of the curated literature on eco-driving studies}\label{tbl1}
		\centering
		\begin{adjustbox}{width=\columnwidth,center}
			\begin{threeparttable}
				\begin{tabular}{>{\centering\arraybackslash}p{0.13\textwidth}>{\centering\arraybackslash}p{0.15\textwidth}>{\centering}p{0.32\textwidth}>{\centering}p{0.1\textwidth}>{\centering}p{0.4\textwidth}>{\centering\arraybackslash}p{0.1\textwidth}}
					\toprule
					Authors     & Type & Objective of the study & Setup & Major aspects and findings & Fuel/Energy savings  \\
					\midrule
					\citet{Beusen2009} & Eco-driving course  & Study the long-term benefits of the eco-driving training from the data collected for several months & Field trials (P=10)   & Immediate improvement in fuel consumption was observed, but some drivers tend to fall back to their original driving style & 5.8\% \\     \\
					
					\citet{Daun2013} & Predictive eco-driving assistance system  & Evaluating the efficacy of a predictive advisory system for heavy commercial vehicles  & Driving simulator (P=40) & Instructing to drive fuel-efficiently and indication of advisory messages to drivers helped in energy consumption reductions. However, the energy consumption rose after immediate tests without the assistance  & 6.6 - 12.2\% \\ 		\\
					
					\citet{Vagg2013} & Retrofit driver advisory system & Design and evaluation of the effectiveness of an real-time feedback system that encourages eco-driving in light commercial vehicles  & Field trials (P=15) & Using the assistance system has lowered the rate of accelerations and early upshifting of the gears, thus resulting in reductions in the energy consumption  & 7.6\% \\ 		\\
					
					\citet{Staubach2014} & Eco-driving support system  & Evaluate an eco-driving support system in urban and rural environments while communicating with TL signals and traffic signs  & Driving simulator (P=30) & Fuel-optimal gear shifting indication and green-wave speed suggestion enabled coasting behaviors and reduced stops at the TL signals & 15.9-18.4\% \\ \\
					
					\citet{Zhao2015} & Eco-driving support system  & Investigate an eco-driving support system that provides both static and dynamic feedback to help drivers in reducing emissions and fuel consumption & Driving simulator (P=22) & Suggestions are provided to avoid rapid acceleration and deceleration events, limit engine RPM, avoid unstable speeds on freeways and idling for a longer time & 3.43-5.45\% \\ 		\\
					
					\citet{Fors2015} & Eco-driving support system  & Investigate various types of in-vehicle eco-driving feedback systems that advise on fuel-efficient driving   & Driving simulator (P=24) & Drivers showed a positive attitude towards the eco-driving support systems, however fuel-efficiency was not investigated in their work & - \\ 		\\
					
					\citet{Sullman2015} & Eco-driving training  & Study whether the professional bus drivers trained with the eco-driving techniques on a simulator would be able to implement these learnings in daily life & Driving simulator (P=29) & Points out that the simulator-based eco-driving training has significant fuel-saving benefits in real-world driving as well    & 11.6-16.9\%  \\ \\				
					
					\citet{Jamson2015} & Eco-driving feedback system  & Evaluate both visual and haptic feedback systems for the uphill and downhill driving scenarios & Driving simulator (P=22) & Continuous real-time visual feedback to the drivers was proven to be more effective than the haptic-based system 
					& - \\ 		\\
					
					\citet{Beloufa2017} & Interactive guidance system  & Measure the contribution of a real-time interactive guidance system to teach eco-driving using immersive simulation  & Driving simulator (P=72) & Teaching eco-driving behavior through instructional videos and interactive guidance system had positive impact on fuel savings and CO2 emission reduction  & 7.42-12.38\% \\ 		\\
					
					\citet{Ayyildiz2017} & Eco-driving training & Evaluate the benefits of an eco-driving based advanced telematics platform in reducing the fuel consumption in freight transport & Field trials (P=25) & The real-time information and metrics on time lost, grade work, normalized braking and acceleration index for the trip helped the drivers to obtain energy savings & 5.5\% \\ 		\\

					\citet{Ruben2021} & Eco-driving training  & Investigate the factors that can reinforce the benefits from an eco-driving theory course in the freight segment & Field trials (P=14) & To strengthen the benefits of an eco-driving course in long term, follow-up evaluations through performance reports and non-monetary rewards were found to be significant   & 7.5-9\% \\ 		\\
					
					\citet{Miotti2021} & Eco-driving heuristics  & Investigate on how modifying the driving style using heuristics could lead to a reduction in the energy consumption  & Simulation & Modify the representative driving cycles into eco-driving speed profiles by limiting maximum travel speed, intensity of acceleration and braking at high speeds and encouraging coasting behavior & 6\% \\ 		\\
					
					\citet{Gonzalez2021} & On-board driving assistance device  & Monitor the fuel savings in waste-collection fleet by implementing a real-time eco-driving feedback to the drivers   & Field trials (P=67) & Assisting drivers to minimize excessive idling, over-revving, harsh braking and over-speeding in real-time using visual green/red lights and auditory signals  & 7.45\%      \\ 
					\bottomrule                                             
				\end{tabular} 	
				\begin{tablenotes}
					\item[]{P = Number of participants}
				\end{tablenotes}		
			\end{threeparttable}
		\end{adjustbox}
	\end{table} 	
\subsection{Related Work on User Acceptance}
To evaluate the acceptance of a new technology, researchers in the past have used theories of human behaviour \citep{Rahman2017} to understand the intention towards usage of a particular technology. Several acceptance models are available in the literature that aim to identify the influencing factors that contribute to the user's behavior of accepting a new technology. The most widely used models are the technology acceptance model (TAM) \citep{Davis1985, Davis1989}, theory of planned behaviour (TPB) \citep{Ajzen1991}, and the unified theory of acceptance and use of technology (UTAUT) \citep{Venkatesh2003, Venkatesh2012}. These models measure the acceptance of a technology using the behavioral intention (BI) to use the technology and its actual use. In the initial years, these models were used to evaluate the user acceptance in various computer-based information systems \citep{Igbaria1993}. 

Previous studies on the user acceptance in advanced driver assistance systems (ADAS) adopted the existing theories and models to evaluate the factors that contribute to the intention to use  state-of-the-art ADAS systems such as adaptive cruise control, lane keeping assist and lane departure warning systems \citep{Rahman2017,Lyu2019,Voinea2020}. \citet{Rahman2017} studied TAM, TPB, and UTAUT for modeling driver acceptance in ADAS, and gathered subjective data through both driving simulator and online survey approaches. Their findings indicated that all three models were able to explain the driver acceptance in terms of behavioral intention (BI) to use ADAS. Moreover, TAM was found to be better at predicting BI, and TPB was regarded as the second best. It also showed that the moderating factors of UTAUT were not significant in predicting BI. Furthermore, TAM was used in \citet{Lyu2019} to evaluate the effectiveness of lane departure warning (LDW) and forward collision warning (FCW) assistance systems through field operation tests. Their study revealed that the drivers' acceptance of FCW was much higher than LDW. They also stated that the road type and driving experience have a significant influence on driving behaviors. \citet{Voinea2020} utilized TAM to assess the driver acceptance of a smartphone-based ADAS that combined navigation and collision warning features with an objective to improve driving safety. The results of their experiment showed that TAM was able to explain significant variance in BI, thus user acceptance as well. The study revealed that the main factors that influenced the users in accepting the system were attitude and perceived usefulness. The limitation in their study is the relatively small sample size used of 23 participants that may limit the generalizability of the findings. Furthermore, the effect of moderating factors such as varying age, gender and driving experience on the acceptance of ADAS was investigated in previous works \citep{Son2015,Lyu2019,Gunthner2021}.

The literature is rich in studies that evaluated the user acceptance on simplified eco-driving assistance systems for ICE vehicles \citep{Vagg2013,Staubach2014,Henzler2015}. The authors in \citet{Staubach2014} evaluated an eco-driving support system that suggested optimal gear-shifting indication and acceleration/deceleration behaviors in less complex scenarios. The acceptance questionnaire was designed based on perceived ease of use, perceived usefulness and behavioral intention to use, and were evaluated on a likert scale. Moreover, \cite{VanDerLaan1997} scale was used to determine the attitude towards using the system. Moreover, \citep{Henzler2015} used UTAUT to evaluate the driver acceptance of various eco-driving feedback systems such as haptic, graphical user interface (GUI) and series HMI in heavy duty vehicles. Very limited studies in the past have evaluated the user acceptance on advanced eco-driving assistance systems, specifically with a focus on battery electric vehicles. To determine the constructs that influence drivers’ intentions to accept connected eco-driving technology for both ICEs and BEVs, \citep{Lin2022} studied TAM, TPB and goal framing. Their results revealed that the	perceived behavioral control (PBC) in TPB showed the most substantial impact on intention to use eco-driving technology. Furthermore, results showed that the BEV drivers possessed a greater understanding of eco-driving in comparison to ICE drivers. The limitation in their study is that the participants were presented with the eco-driving system using video demonstrations and could not experience the system in real-world or in a driving simulator. In our study, the main objective is to evaluate the factors that influence the behavior intention to use a predictive eco-driving assistance system (pEDAS) by adopting existing TAM and TPB models, with a focus towards battery electric vehicles.

	\section{Methodology}
	\label{sec:methodology}
	\subsection{Layout of the Eco-Driving Assistance System}
	\label{sec:layout}
\begin{figure}
		\centering
		\includegraphics[width=0.7\linewidth]{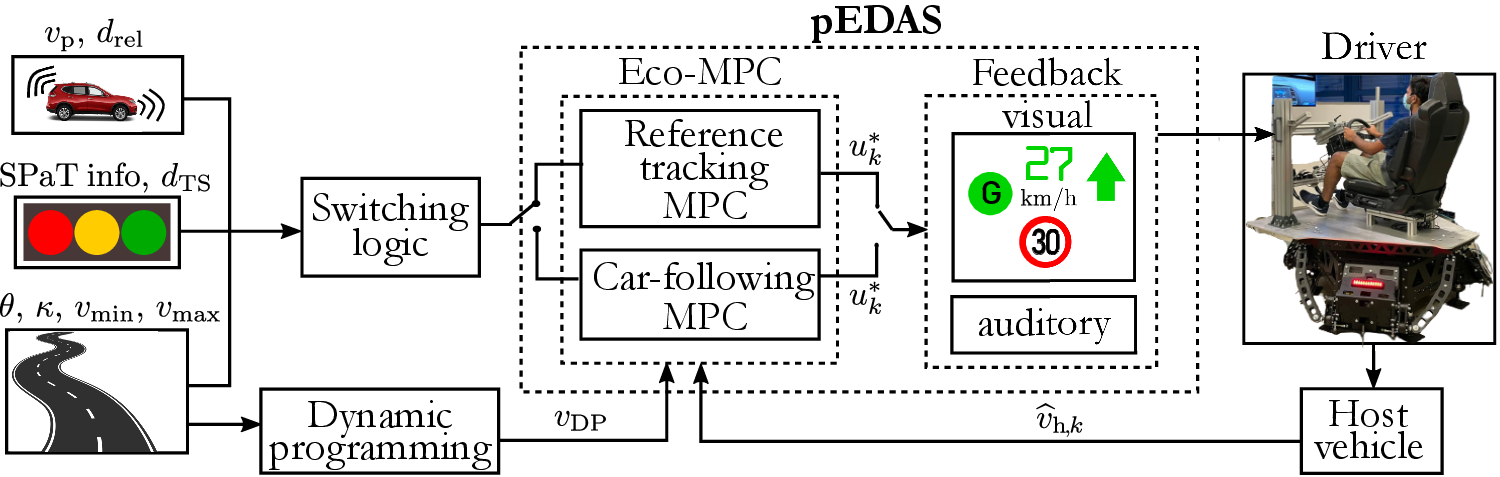}
		\caption{Layout of the proposed predictive eco-driving assistance system}
		\label{fig:EDAS_Layout}
	\end{figure}

	The layout of the proposed predictive eco-driving assistance system (pEDAS) is presented in Figure~\ref{fig:EDAS_Layout}. pEDAS in this work is a combination of two linear model predictive controllers (MPCs), namely reference-tracking MPC and car-following MPC, along with an eco-driving feedback system. pEDAS uses the information gathered from onboard sensors (radar or lidar), for instance the relative distance $d_\text{rel}$ to the preceding vehicle and its velocity $v_\text{p}$, SPaT information of the upcoming TL signals from the road side units (RSUs), and relative distance $d_\text{ITS}$ between the host vehicle and the upcoming signalized intersections or stop signs in the driving route, geographical information of the driving route such as elevation $\theta$, road curvature $\kappa$, road minimum $v_\text{min}$ and maximum $v_\text{max}$ speed limits and the optimal speed profile $v_\text{DP}$ calculated offline using dynamic programming and finally, the measured host vehicle velocity $\widehat{v}_\text{h}$. The switching logic primarily uses the inter-vehicle distance $d_\text{rel}$ between the host vehicle and the preceding vehicle as the basis for choosing a suitable controller. In a freeway scenario, when a preceding vehicle is not in the sensor range of the host vehicle ($d_\text{rel}$ > \unit[100]{m}), the reference-tracking MPC is engaged. Conversely, when the preceding vehicle is within the sensor range, the car-following MPC is selected. The energy-optimal control sequence $u^{*}_{k}$ is obtained from either of these controllers at each sample time step $k$, which is in in turn used to compute an optimal speed for the driver. Furthermore, the optimal speed is communicated to the driver through an eco-driving feedback system in a visual and auditory manner. Eventually, the driver tracks these continuous eco-driving speed recommendations to drive the host vehicle energy-efficiently. 
	\subsection{Reference Speed Calculation}
	\label{sec:green_wave_velcoity_calculation}
\begin{figure*}[t!]
	\centering
	\begin{subfigure}[t]{0.5\textwidth}
		\centering
		\includegraphics[width=1.0\textwidth]{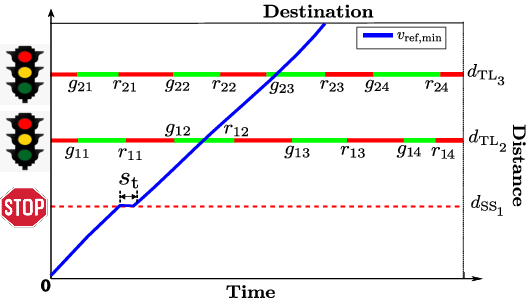}
		\caption{Exemplary green-wave optimal speed profile}
		\label{fig:greenwave_profile}
	\end{subfigure}%
	\begin{subfigure}[t]{0.5\textwidth}
		\centering
		\includegraphics[width=0.8\textwidth]{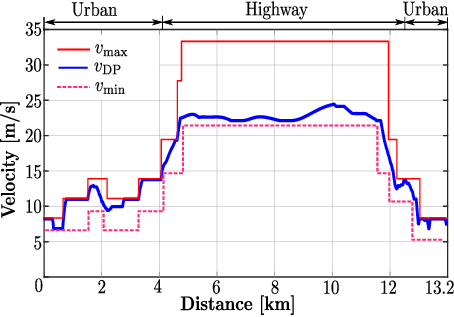}
		\caption{Speed profile generated using dynamic programming}
		\label{fig:DP_profile}
	\end{subfigure}
	\caption{Reference speed profile generation}
\end{figure*}

	The goal of the reference-tracking MPC in this work is to track a pre-determined reference speed in a freeway scenario, which is generated either from a green-wave algorithm or a dynamic programming algorithm. A green-wave algorithm \citep{Asadi2011,Chada2020} as given in (\ref{eqn:intersection}) is used to calculate a green-wave reference speed for the host vehicle to cross the upcoming TL signals at a green phase. It is assumed in this work that the signal phasing and timing (SPaT) information of the upcoming TL signals is communicated to the host vehicle in real-time through RSUs. 
	\begin{align}
	[v_\text{ref,min} , v_\text{ref,max}] =	\left[\frac{d_{\text{TL}_{m}}}{r_{mn\ }} ,\frac{d_{\text{TL}_{m}} }{g_{mn\ }}\right] \cap [ v_\text{min} ,v_\text{max}] \ 
	\label{eqn:intersection}
	\end{align}	
In (\ref{eqn:intersection}), $d_{\text{TL}_{m}}$ represents the distance to the $m^{th}$ TL signal and $r_{mn}$ represents the start time of the $n^{th}$ red phase of the $m^{th}$ TL signal. Similarly, $g_{mn}$ represents the start time of the $n^{th}$ green phase of the $m^{th}$ TL signal. Furthermore, $v_\text{min}$ and $v_\text{max}$ are the minimum and maximum road speed limits respectively. From the possible green-wave intersection range obtained using (\ref{eqn:intersection}), $v_\text{ref,min}$ is chosen as the reference velocity, because travelling at lower velocities is generally more energy-efficient than at higher velocities. To explain the green-wave optimal speed profile in a simplified way, an exemplary scenario using a space-time diagram is illustrated in Figure~\ref{fig:greenwave_profile}. The driving route in the exemplary scenario consists of a stop sign $\text{SS}_{\{1\}}$ and two signalized intersections $\text{TL}_{\{2,3\}}$. It is to be noted that the green-wave intersection using (\ref{eqn:intersection}) is specifically calculated for the upcoming TL signals alone and the stop signs are exempted. This is to ensure that the host vehicle performs a compulsory halt at each stop sign for a designated dwell time period $s_\text{t}$, considered as \unit[3]{s} in this work. To avoid the scenarios in which the green-wave algorithm suggests drivers to travel at very low speeds that might cause inconvenience to the following vehicles, a minimum road speed limit $v_\text{min}$ is introduced in (\ref{eqn:intersection}) as depicted in Figure~\ref{fig:DP_profile}. If a green-wave intersection range is not found using (\ref{eqn:intersection}), it infers that stopping at the next TL signal is unavoidable. In such a case, the host vehicle is advised to travel with an optimal speed calculated offline using dynamic programming (DP) until reaching the TL signals to stop at a red phase. The DP cost function in this study considers the road elevation, speed limit constraints and a trade-off between trip time and minimizing energy consumption. Further details about the modeling of the cost function and constraints can be found in \citet{Lin2014,Weissmann2018}. The maximum allowable road speed considering road curvature $v_\text{max}$, the minimum allowable road speed $v_\text{min}$ and the reference DP speed profile $v_\text{DP}$ computed at the beginning of the trip are illustrated in Figure~\ref{fig:DP_profile}. It is to be mentioned here that the weightage in the DP algorithm was adjusted to favor more energy savings over the smaller trip time, thus resulting in lower mean speeds on highway segment as depicted in Figure~\ref{fig:DP_profile}.
	\subsection{System Dynamics}
	\label{sec:System_Dynamics}
	The longitudinal vehicle dynamics of the host vehicle is described by
	\begin{align}
	\frac{dv_\text{h}}{dt} = \frac{1}{m_\text{eq}}(F_\text{t} - F_\text{b} - \underbrace{F_\text{a} - F_\text{r} - F_\text{g}}_{F_\text{res}})
	\label{eqn:Acceleration_1}
	\end{align} 
	where $v_\text{h}$ is the host vehicle velocity, $m_\text{eq}$ is the equivalent mass which is the sum of vehicle weight, rotational equivalent masses, driver and cargo weight,  $F_\text{t}$ is the traction force, $F_\text{b}$ is the braking force, $F_\text{res}$ is the combination of aerodynamic resistance $F_\text{a} = \frac{1}{2}\rho A_\text{f} c_\text{w} v_\text{h}^2$, rolling resistance $F_\text{r}=c_\text{r}m_\text{v}g \cos\theta$ and gradient resistance $F_\text{g} = m_\text{v}g \sin\theta$. To handle the nonlinearity occurring due to the term $v_\text{h}^2$ in $F_\text{a}$, an approximation of the aerodynamic resistance $\bar{F_\text{a}} \approx \frac{1}{2}\rho A_\text{f} c_\text{w} (p_{1}v_\text{h}+p_\text{2})$ is considered in this work as illustrated in Figure~\ref{fig:app_aero}. Here, $c_\text{w}$ is the drag coefficient, $A_\text{f}$ is the frontal cross-sectional area of the vehicle, $\rho$ is the density of the air, $p_\text{1}$ and $p_\text{2}$  are the coefficients obtained through line fitting, $c_\text{r}$ is the coefficient of rolling resistance, $m_\text{v}$ is the host vehicle weight, $g$ is the gravitational acceleration and $\theta$ is the gradient angle. 	 
	
	The host vehicle velocity at control step $k+1$ after discretization using the timestep $\Delta T$ (chosen as \unit[0.2]{s}) can be described by (\ref{eq:discretized_1}). The relative distance $d_\text{ITS}$ between the host vehicle and the upcoming TL signal or a stop sign at control step $k+1$ is updated using (\ref{eq:discretized_2}), where $\text{ITS}=\{\text{TL},\text{SS}\}$. Similarly, while performing car-following, the relative distance $d_\text{rel}$ between the host vehicle and the preceding vehicle at control step $k+1$ can be calculated using (\ref{eq:discretized_3}), where $v_{\text{p},k}$ is the velocity of the preceding vehicle. 
	\begin{subequations}
		\begin{flalign} 
		\label{eq:discretized_1}
		& v_{\text{h},k+1}= v_{\text{h},k} + \frac{\Delta T }{m_\text{eq}}( F_{\text{t},k}- F_{\text{b},k} - F_{\text{res},k}) \\ \label{eq:discretized_2}
		& d_{\text{ITS},k+1}=d_{\text{ITS},k} - \Delta T \left(\frac{v_{\text{h},k}+v_{\text{h},k+1}}{2}\right) \\ \label{eq:discretized_3} & 
		d_{\text{rel},k+1}=d_{\text{rel},k}+ \Delta T \left( \frac{v_{\text{p},k}+v_{\text{p},k+1}}{2} -\frac{v_{\text{h},k}+v_{\text{h},k+1}}{2}\right)
		\end{flalign}
		\label{eq:discretized_4}  
	\end{subequations}
	The chosen host vehicle in this work is a Nissan Leaf \citep{Wager2016} BEV, whose accurate vehicle and battery models are obtained from \citet{Lin2014}. A half-map approximation of the BEV power consumption map as discussed in our previous work \citep{Chada2020} is used in this study. The BEV power consumption is defined as a function of velocity $v_\text{h}$ and traction force $F_\text{t}$ using (\ref{eq:power_consumption}), where $a_{00}$, $a_{10}$, $a_{01}$, $a_{11}$, $a_{20}$, $a_{02}$ are the coefficients obtained through a second-order polynomial approximation, i.e.
	\begin{equation} 
	\label{eq:power_consumption}
	P(v_\text{h}, F_\text{t}) = a_{00}+a_{10}v_\text{h}+a_{01}F_\text{t}+a_{11}v_\text{h}F_\text{t} +a_{20}v_\text{h}^2+a_{02}F_\text{t}^2 
	\end{equation} 
	\subsection{Constraints}
	To ensure effective implementation of the proposed controllers, several constraints must be included. For instance, to incorporate the physical limitations of the electric motor, the host vehicle traction force $F_\text{t}$ is chosen to be non-negative and is constrained by the maximum traction force $F_\text{t,max}\approx p_\text{3}v_{\text{h},k}+p_\text{4}$ as given in (\ref{eq:constraints_1}), where $p_{3}$ and $p_{4}$ are parameters obtained through line approximation \citep{Weissmann2018}. Additionally, the physical limitations on the vehicle in terms of braking force $F_\text{b}$ and the velocity $v_\text{h}$ are constrained by (\ref{eq:constraints_3}) and (\ref{eq:constraints_4}) respectively.
	\begin{subequations}
		\begin{flalign} 
		\label{eq:constraints_1}
		& 0 \leq F_{\text{t},k} \leq  F_\text{t,max} \\	\label{eq:constraints_3}   
		& 0 \leq F_{\text{b},k} \leq  F_\text{b,max} \\	\label{eq:constraints_4} 
		& v_{\text{min},k} \leq v_{\text{h},k} \leq  \text{min}(v_{\text{max},k},v_{\text{DP},k})
		\end{flalign}
		\label{eq:physical_limits}  
	\end{subequations}
	As defined in the constant time headway policy \citep{Li2009}, the host vehicle must maintain a safe as well as desired distance to the preceding vehicle while performing car-following. To keep a safe distance $d_\text{s}$ (as illustrated in Figure~\ref{fig:with_preceding_car}) to the preceding vehicle, a hard constraint is imposed using (\ref{eq:d_safe}), where $h_\text{s}$ is the minimum constant time headway and $d_\text{min}$ is the minimum safety distance during standstill. Similarly, to motivate the host vehicle to maintain a desired inter-vehicle distance to a preceding vehicle $d_\text{c}$, a soft constraint is defined using (\ref{eq:d_comfort}), where $h_\text{c}$ is the maximum constant time headway and $\varepsilon_\text{3}$ is a slack variable (as illustrated in Figure~\ref{fig:with_preceding_car}).
	\begin{subequations}
		\begin{flalign} 
		\label{eq:d_safe}
		& d_{\text{rel},k} \geq \underbrace{d_\text{min} + h_\text{s} v_{\text{h},k}}_{d_{\text{s},k}} \\ \label{eq:d_comfort} 
		& d_{\text{rel},k} \leq \underbrace{ d_\text{min} + h_\text{c} v_{\text{h},k}}_{d_{\text{c},k}} + \varepsilon_{\text{3},k}
		\end{flalign}
		\label{eq:d_safe_comfort}  
	\end{subequations}
	Moreover, to minimize jerks and increase the driving comfort, a soft constraint on the change in vehicle traction force between two successive steps is enforced using (\ref{eq:constraints_5}) and (\ref{eq:constraints_6}), where  $\Delta F_\text{t,max}$ is the maximum allowable variation limit and $\varepsilon_\text{2}$ is a slack variable.  
	\begin{subequations}
		\begin{flalign} 
		\label{eq:constraints_5}
		& F_{\text{t},k}- F_{\text{t},k+1} - \varepsilon_{\text{2},k} \leq \Delta F_\text{t,max} \\	\label{eq:constraints_6}
		& F_{\text{t},k+1} - F_{\text{t},k} - \varepsilon_{\text{2},k} \leq \Delta F_\text{t,max}
		\end{flalign}
		\label{eq:traction_force}  
	\end{subequations}
	For the host vehicle to avoid crossing the stop signs or TL signals at red, a safety constraint is enforced using (\ref{eq:constraints_7}). Moreover, to avoid undesirable scenarios where the host vehicle might stop far away from the stop sign or the TL signals, a soft constraint is defined using (\ref{eq:constraints_8}), where $d_\text{s,ITS}$ is the distance from the stop line to the stop sign or TL signal and $\varepsilon_\text{1}$ is a slack variable (as illustrated in Figure~\ref{fig:driving_scenarios}).
	\begin{subequations}
		\begin{flalign} 
		\label{eq:constraints_7}
		& d_{\text{ITS},k} \geq 0 \\	\label{eq:constraints_8}   	
		& d_{\text{ITS},k} - \varepsilon_{\text{1},k} \leq d_\text{s,ITS}
		\end{flalign}
		\label{eq:TL_distance}  
	\end{subequations}
	\subsection{Problem Formulation}
	\label{sec:problem_formulation}
	\begin{figure*}[t!]
		\centering
		\begin{subfigure}[t]{0.5\textwidth}
			\centering
			\includegraphics[width=0.8\textwidth]{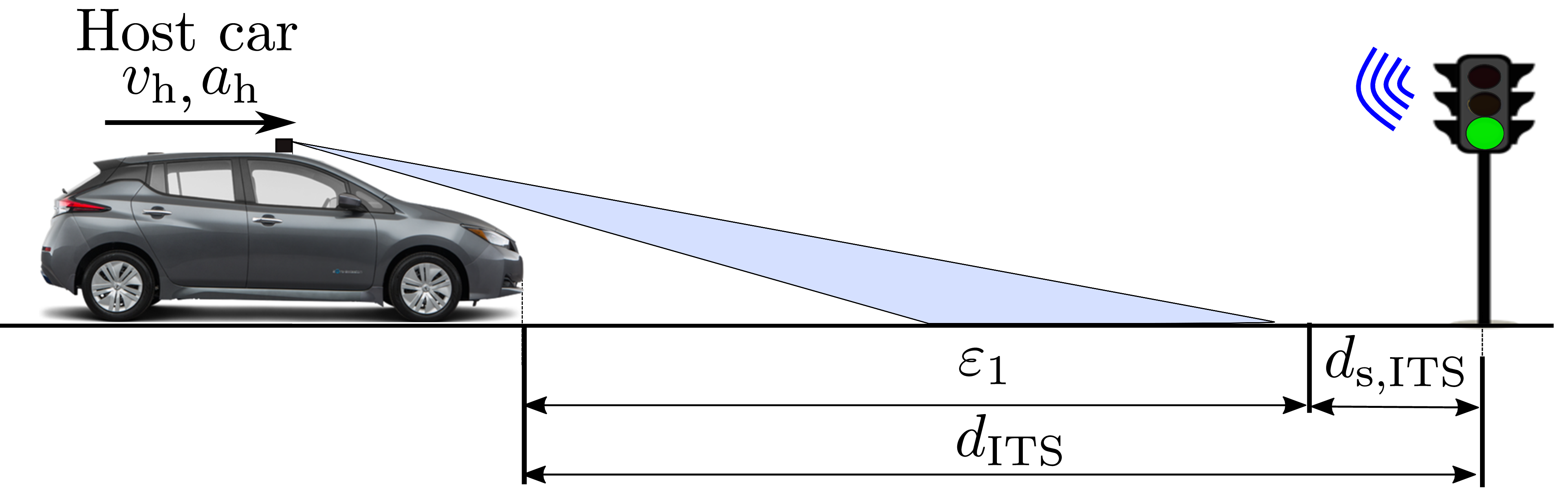}
			\caption{Freeway scenario with no preceding vehicle in range}
			\label{fig:no_preceding_car}
		\end{subfigure}%
		~ 
		\begin{subfigure}[t]{0.5\textwidth}
			\centering
			\includegraphics[width=0.9\textwidth]{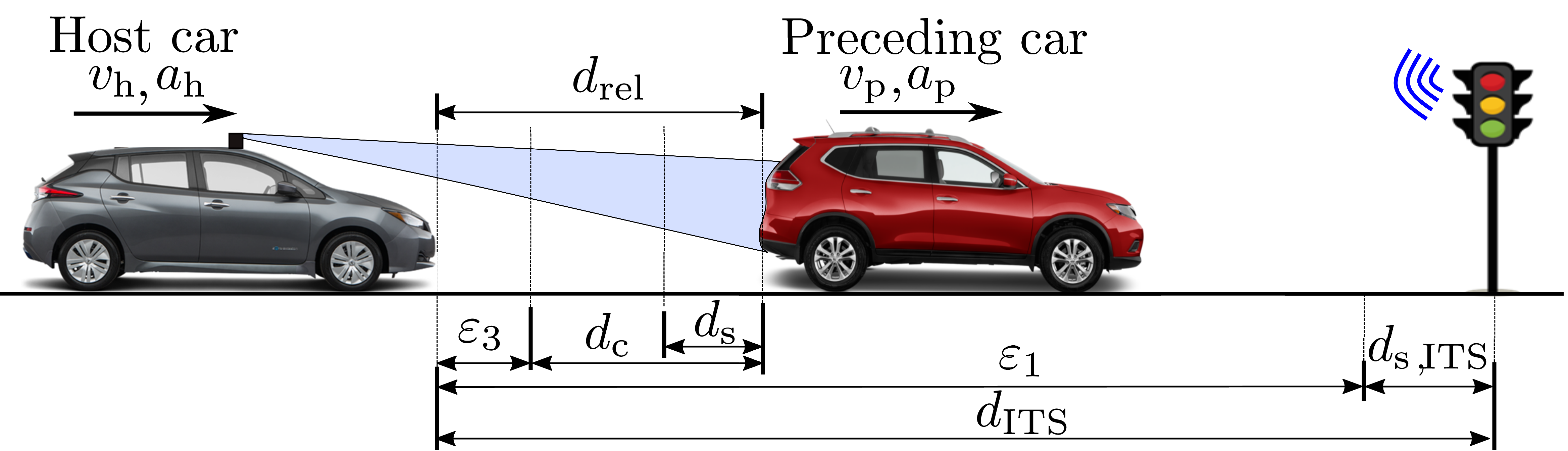}
			\caption{Car-following scenario with preceding car in range}
			\label{fig:with_preceding_car}
		\end{subfigure}
		\caption{Commonly occurring traffic scenarios for a host vehicle}
		\label{fig:driving_scenarios}
	\end{figure*}
	In this section, the problem formulations for the proposed MPCs are introduced. Firstly, the quadratic cost function for the reference-tracking MPC for a freeway scenario (Figure~\ref{fig:no_preceding_car}) is defined using (\ref{eq:Green_wave_cost_function}). The cost function is minimized over a control sequence  $u^{*}_{k}= [F_{\text{t},k},F_{\text{b},k},\varepsilon_{\text{1},k},\varepsilon_{\text{2},k}]^\intercal$ along the prediction horizon $N=20$ steps chosen as a trade-off between real-time computation and energy savings. The first term in (\ref{eq:Green_wave_cost_function}) minimizes the BEV power consumption as a function of $v_{h,k}$ and $F_{t,k}$ (\ref{eq:power_consumption}). The second term includes a penalty for the deviation of the host vehicle velocity $v_{\text{h},k}$ from the reference velocity $v_{\text{ref},k}$. The third term exists to penalize the excessive braking force $F_{\text{b},k}$.In case, a green-wave intersection (\ref{eqn:intersection}) is not found, the fourth term ensures a penalty on the slack variable $\varepsilon_{1,k}$, to motivate the host vehicle to reach as close as possible to the stop line $d_\text{s,ITS}$ at a stop sign or red TL signal. The final term enforces a penalty on the slack variable $\varepsilon_{2,k}$, which ensures to reduce excessive jerks and improve the driving comfort. Moreover, $\zeta_{1}$, $\zeta_{2}$, $\zeta_{3}$ and $\zeta_{4}$ are the respective weights of the aforementioned terms. 
	\begin{subequations}
		\begin{flalign} 
		\label{eq:Green_wave_cost_function}
		& \min_{\textit{\textbf{$F_{t,k}$}},\textbf{$F_{b,k}$},{\varepsilon}_{1,k},{\varepsilon}_{2,k}}
		\sum_{k=0}^{N-1} P(v_{\text{h},k}, F_{\text{t},k}) + \zeta_{1}\Vert v_{\text{h},k}-v_{\text{ref},k}\Vert ^{2} + \zeta_{2}\Vert F_{\text{b},k}\Vert ^{2} + \zeta_{3}\Vert \varepsilon_{1,k}\Vert ^{2} + \zeta_{4}\Vert \varepsilon_{2,k}\Vert ^{2} \\ 
		& \text{s.t.} \ \ (\text{\ref{eq:discretized_1})}, (\text{\ref{eq:discretized_2}}), (\text{\ref{eq:constraints_1}}), (\text{\ref{eq:constraints_3}}), (\text{\ref{eq:constraints_4}}), (\text{\ref{eq:constraints_5}}), (\text{\ref{eq:constraints_6}}), (\text{\ref{eq:constraints_7}}), (\ref{eq:constraints_8}) \nonumber
		\end{flalign}
		\label{eq:Online_Computation_25}  
	\end{subequations}
	The quadratic cost function for the car-following MPC is described using (\ref{eq:MPC_Cost_Function}) and is minimized over a control sequence  $u^{*}_{k}= [F_{\text{t},k},F_{\text{b},k},\varepsilon_{\text{1},k},\varepsilon_{\text{2},k},\varepsilon_{\text{3},k}]^\intercal$ along the prediction horizon $N$ similar to (\ref{eq:Green_wave_cost_function}). 
		\begin{subequations}
		\begin{flalign} 
		\label{eq:MPC_Cost_Function}
		&\min_{\textit{\textbf{$F_{t,k}$}},\textbf{$F_{b,k}$},{\varepsilon}_{1,k},{\varepsilon}_{2,k},{\varepsilon}_{3,k}}\sum_{k=0}^{N-1} P(v_{\text{h},k}, F_{\text{t},k}) + \zeta_{1}\Vert \varepsilon_{3,k}\Vert ^{2} + \zeta_{2}\Vert F_{\text{b},k}\Vert ^{2} + \zeta_{3}\Vert \varepsilon_{1,k}\Vert ^{2} + \zeta_{4}\Vert \varepsilon_{2,k}\Vert ^{2} \\ \label{eq:Online_Computation_4} 
		& \text{s.t.} \ \ (\text{\ref{eq:discretized_1})}, (\text{\ref{eq:discretized_2}}), (\text{\ref{eq:discretized_3}}),  (\text{\ref{eq:constraints_1}}), (\text{\ref{eq:constraints_3}}), (\text{\ref{eq:constraints_4}}), (\text{\ref{eq:d_safe}}), (\text{\ref{eq:d_comfort}}), (\text{\ref{eq:constraints_5}}), (\text{\ref{eq:constraints_6}}), (\text{\ref{eq:constraints_7}}),  (\text{\ref{eq:constraints_8}})\nonumber
		\end{flalign}
		\label{eq:Online_Computation}  
	\end{subequations}	
	The only difference is that instead of penalizing the deviation from reference velocity, the slack variable $\varepsilon_{3,k}$ is penalized in the second term to motivate the host vehicle to stay within a desired inter-vehicle distance to the preceding vehicle (Figure~\ref{fig:with_preceding_car}). It is to be noted that the future velocities of the preceding vehicle are assumed to be constant within the prediction horizon in this work, thus corresponding to a frozen-time model predictive controller (FTMPC) \citep{Ripaccioli2010}. Both (\ref{eq:Online_Computation_25}) and (\ref{eq:Online_Computation}) were solved using a quadprog solver available in MATLAB R2019b \citep{Abebe2007}. The performance of the MPC controllers described in (9a) and (10a) has been extensively discussed in our previous contribution \citep{Chada2020}. When evaluated under realistic urban driving scenarios, these controllers showcased great switching ability, energy-saving benefits, and online implementation capabilities as well. 
	
	It is to be mentioned that the non-negative optimal braking force $F_{\text{b},k}$ calculated by solving the optimization problems (\ref{eq:Green_wave_cost_function}) and (\ref{eq:MPC_Cost_Function}) consists of both the regenerative and friction braking forces. If $F_{\text{b},k} \leq \lvert -F_\text{t,max} \rvert$, then $F_{\text{b},k}$ is the regenerative braking force. If $F_{\text{b},k} > \lvert -F_\text{t,max} \rvert$, then $\lvert -F_\text{t,max} \rvert$ is the regenerative braking force, and $F_{\text{b},k} - \lvert -F_\text{t,max} \rvert$ is the friction braking force. In BEVs, deceleration resulting due to excessive braking can cause in loss of energy as well, as it contributes to friction braking in which the kinetic energy is converted into heat and then dissipated into the environment. Therefore, with the term $F_{\text{b},k}^{2}$ in both problem formulations (\ref{eq:Green_wave_cost_function}) and (\ref{eq:MPC_Cost_Function}), excessive braking force is penalized.    
	\subsection{MPC Infeasibility}
	\label{sec:driver_disturbance}
\begin{figure*}[t!]
	\centering
	\begin{subfigure}[t]{0.5\textwidth}
		\centering
		\includegraphics[width=0.5\textwidth]{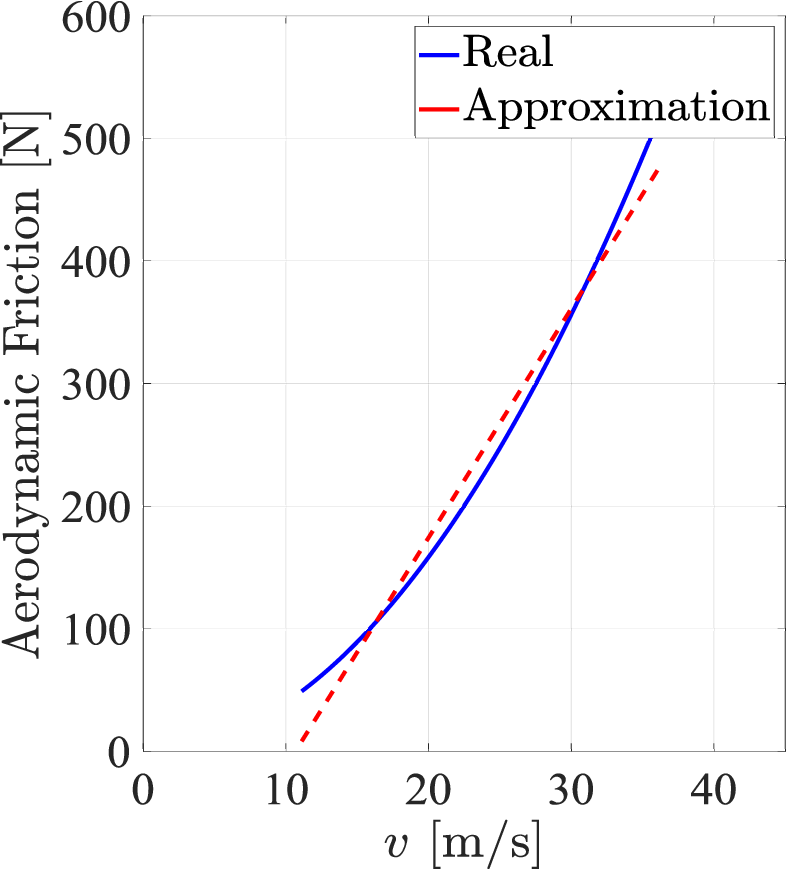}
		\caption{Approximation of aerodynamic friction}
		\label{fig:app_aero}
	\end{subfigure}%
	\begin{subfigure}[t]{0.5\textwidth}
	\centering
	\includegraphics[width=0.5\textwidth]{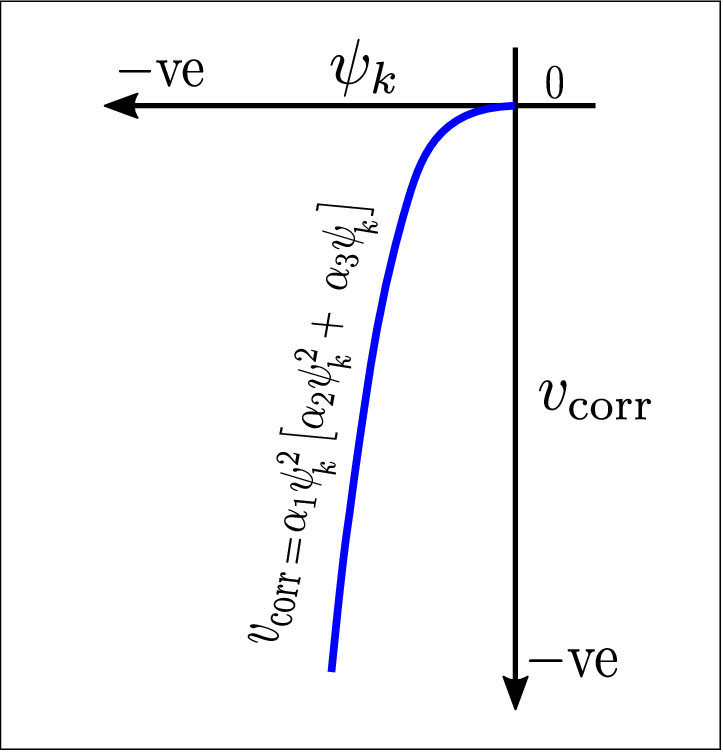}
	\caption{Speed correction function}
	\label{fig:speed_corr}
	\end{subfigure}
	\caption{Functions for aerodynamic friction and speed correction}
\end{figure*}

It is important to mention here that an inefficient tracking of the optimal speed due to the perception and reaction delay of the driver may act as an external disturbance to the MPC. The resulting speed tracking errors may lead to a violation of safety critical constraints, due to which the MPC may not provide a feasible solution. For instance, while car-following if the driver violates the safe distance $d_s$ to a preceding vehicle due to speed tracking delay, the MPC infeasibility occurs as the hard constraint (\ref{eq:d_safe}) is not satisfied. Therefore, the velocity state is updated at time step $k+1$ using a correction function $v_\text{corr}$ as given in (\ref{eq:v_correction_2}), where $\alpha_{1}= 0.008$, $\alpha_{2}= -0.05$, $\alpha_{3}=0.4$ and $\psi_k = d_{\text{rel},k} - d_{\text{s},k}$. 
	\begin{equation}
	\label{eq:v_correction_2}
	v_{\text{h},k+1}= v_{\text{h},k} + \underbrace{\alpha_{1}\psi_k^2\left[\alpha_{2} \psi_k^2 +\ \alpha_{3}\psi_k\right]}_{v_\text{corr}}	
	\end{equation}		
	As can be inferred from the Figure~\ref{fig:speed_corr}, the $v_\text{corr}$ decrements parabolically in the third quadrant as $\psi_k$ decreases, i.e. upon violating the safe distance, the more closer the host car gets to the preceding car, the larger is the speed decrement. For simplicity, the parameters in (\ref{eq:v_correction_2}) are kept constant across drivers, and are chosen through repeated trials by evaluating if the drivers are able to decelerate and move outside of the safe region $d_s$ after observing the updated speed $v_{\text{h},k+1}$ via the eco-driving feedback system. It is to be noted that estimating the driver delay and its compensation is outside the scope of this work and has been investigated extensively in our extended work in \cite{Chada2022}. 
	\subsection{Eco-driving Feedback System}
	\label{sec:Ecodriving_Feedback_System}
	\begin{figure}[t]
		\centering
		\setkeys{Gin}{width=\linewidth,height=4cm} 
		\medskip
		\begin{subfigure}{5cm}
			\includegraphics{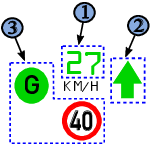}
			\caption{1. Current speed of the host vehicle, 2. Speed advisory using arrow visualization, 3. Additional useful information}
			\label{Fig:a}
		\end{subfigure}
		\hfil
		\begin{subfigure}{5cm}
			\includegraphics{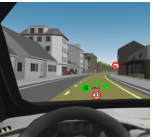}
			\caption{Green-wave speed advisory to cross the immediate TL signal at green phase}
			\label{Fig:b}
		\end{subfigure}
		\hfil
		\begin{subfigure}{5cm}
			\includegraphics{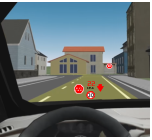}
			\caption{Speed advisory to slow down while approaching a stop sign}
			\label{Fig:c}
		\end{subfigure}
		
		\medskip
		\begin{subfigure}{5cm}
			\includegraphics{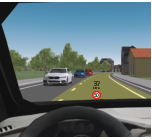}
			\caption{Black digits indicate that current velocity is within the optimal speed band}
			\label{Fig:d}
		\end{subfigure}
		\hfil
		\begin{subfigure}{5cm}
			\includegraphics{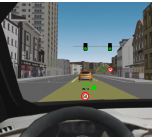}
			\caption{Indication to speed up to maintain a desired inter-vehicle distance to a preceding vehicle}
			\label{Fig:e}
		\end{subfigure}
		\hfil
		\begin{subfigure}{5cm}
			\includegraphics{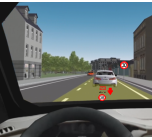}
			\caption{Advisory to slow down to maintain a safe distance to a preceding vehicle}
			\label{Fig:f}
		\end{subfigure}
		
		\medskip
		\begin{subfigure}{5cm}
			\includegraphics{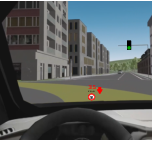}
			\caption{Speed advisory to decelerate and travel at a safe speed while approaching a curved road}
			\label{Fig:g}
		\end{subfigure}
		\hfill
		\begin{subfigure}{5cm}
			\includegraphics{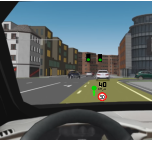}
			\caption{Indication of the remaining time of green traffic light phase}
			\label{Fig:h}
		\end{subfigure}
		\hfill
		\begin{subfigure}{5cm}
			\includegraphics{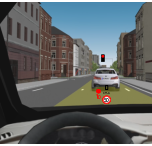}
			\caption{Indication of the remaining time of red traffic light phase}
			\label{Fig:i}
		\end{subfigure}
		\caption{Predictive EDAS suggestions during different driving scenarios in the test route}
		\label{Fig:EDAS_suggestions}
	\end{figure}
	In this work, a continuous eco-driving feedback is provided to the driver in a visual fashion and the driver is alerted during unsafe driving situations through an auditory feedback as well. The visual information is projected onto the windshield of the host vehicle by means of a heads-up display (HUD) and the driver receives the auditory feedback from an in-car speaker. An internal study conducted by the authors explored four different visualization concepts (bar, scale, tachometer, and arrow) for pEDAS. Meanwhile, 22 participants took part in an online survey and rated the visualization concepts through a questionnaire based on \cite{VanDerLaan1997} scales. The designs are evaluated based on the usefulness and satisfaction score obtained from the survey. The results showed that 64\% of participants preferred the arrow type visualization and suggested some improvements to it. Furthermore, the amount of visual information being displayed on the HUD may cause driver distraction and result in an increased cognitive workload. Keeping the aforementioned points in view, an improvised design of arrow visualization with minimal information is presented in this work. An example of the speed advisory system is shown in Figure~\ref{Fig:a} with information being displayed at three regions on the HUD. At the first region, the current speed of the host vehicle is displayed. The second region depicts an arrow indicating the driver to either speed up (pointing upwards and green in color) if the current speed is below the optimal speed or to slow down (pointing downwards and red in color) if above the optimal speed. The arrow length indicates the amount of speed increment or decrement. In the third region, additional information such as traffic light remaining time, green-wave and traffic sign detection is provided to aid the driver to better understand the suggestions given by pEDAS. Moreover, the allowable road speed limit information is displayed below the current speed of the host vehicle. The features of pEDAS are depicted in Figures~\ref{Fig:b}-\ref{Fig:i}. In Figure \ref{Fig:b}, the green-wave speed advisory (Section~\ref{sec:green_wave_velcoity_calculation}) to cross the immediate TL signal at green phase is suggested to the driver. An icon with a letter ’G’ on a green background appears to indicate the driver that a green-wave is possible. pEDAS alerts the driver of a stop or yield sign ahead by displaying a visual icon on the HUD as illustrated in Figure~\ref{Fig:c}. Additionally, it suggests the driver to gradually slow down while approaching the stop/yield sign. This early suggestion helps in minimizing sudden deceleration, thereby jerks as well. Furthermore, in a freeway scenario pEDAS recommends the driver to track a DP optimal speed (Figure~\ref{fig:DP_profile}). Moreover, the suggested optimal speed may vary at each timestamp and therefore maintaining an exact target speed may be difficult for the drivers. Hence, an optimal speed band of ±2 kmph is defined in this work after repeated trials, where the color of the speed digit remains black when the driver reaches this optimal band as shown in Figure~\ref{Fig:d}. In such a case, the drivers are advised to apply a coasting strategy by not pressing the throttle or brake pedals. Moreover, in a car-following scenario, pEDAS suggests the driver to maintain a desired inter-vehicle distance to the preceding vehicle by suggesting to speed-up if the host vehicle violates the comfort distance (Figure~\ref{Fig:e}) and suggests the driver to slow down if the safe distance to the preceding vehicle is about to be violated (Figure~\ref{Fig:f}). In addition, the driver is alerted with a warning tone if the suggestion to slow down is not followed. pEDAS with the help of road curvature information alerts the driver to slow down well ahead of a sharp curve and suggests to take the turn at a safe speed. An example of such a scenario is shown in Figure~\ref{Fig:g}. The next feature displays the remaining time (less than \unit[10]{s}) of the immediate TL signal phase on the HUD when the host vehicle is approaching a signalized intersection. This feature at TL green and red phases is illustrated in Figure~\ref{Fig:h} and \ref{Fig:i} respectively. Since pEDAS uses the SPaT information from the TL infrastructure, displaying the time remaining in a particular signal phase can be helpful for the drivers to prepare for their next actions.
	
	\section{Evaluation}
	\subsection{Setup}
		\begin{figure}
		\centering
		\includegraphics[width=0.9\linewidth]{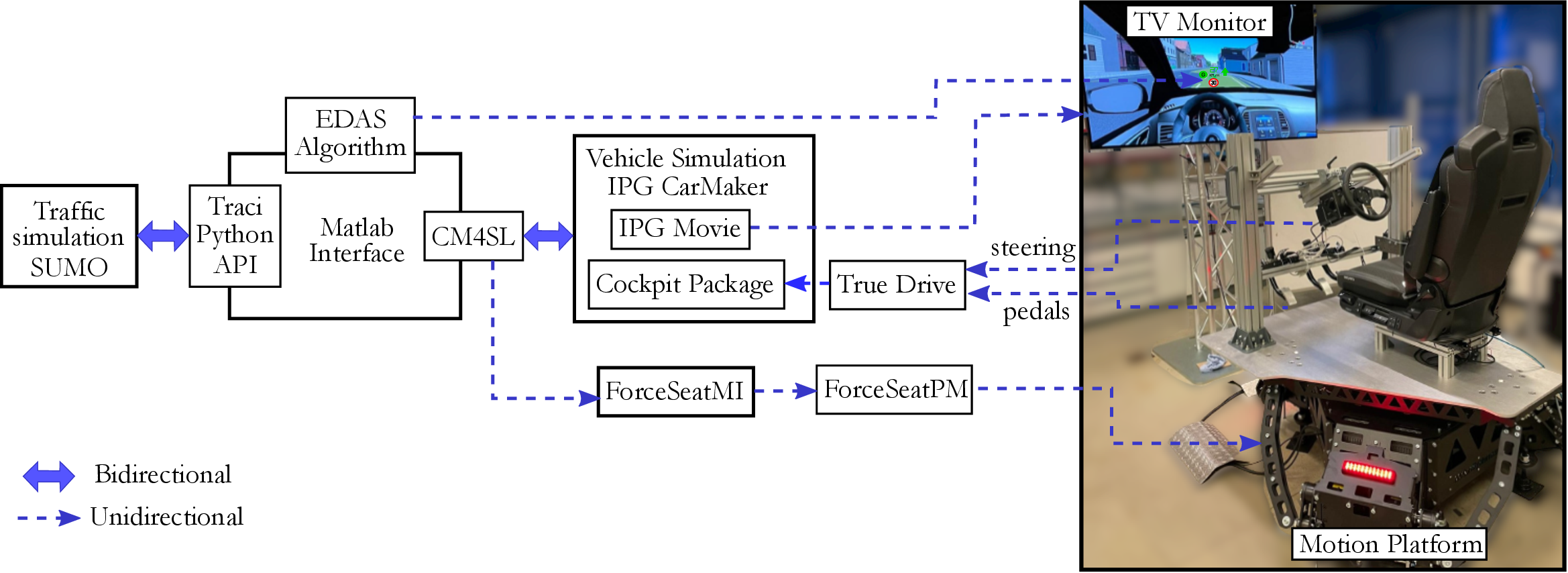}
		\caption{Schematic of the driving simulator setup}
		\label{fig:setup}
	\end{figure}
	The schematic of the driving simulator setup used in this work is illustrated in Figure~\ref{fig:setup}. It consists of a PS-6TM-550 six degree of freedom (6-DoF) motion platform from Motion Systems, a Simucube 2 Pro steering wheel setup, adjustable seat and pedals (throttle and brake). Through the ForSeatMI programming interface and the software engine ForceSeatPM Platform Manager from Motion Systems, the rotational (roll, pitch and yaw) and translational (heave, surge and sway) motions of the dynamic platform can be controlled. The Simucube steering wheel features a direct drive force-feedback motor and realistically simulates the forces from road and the steering linkages. Furthermore, through the Simucube True Drive software, one can easily fine-tune the settings of the steering wheel and pedals to match the feel of a passenger car. In this work, the virtual driving environment was modeled in IPG CarMaker and the driver visualizes the cockpit view on a 65 inch LCD TV monitor through IPG Movie. Furthermore, views from side and rear mirrors were provided so that driver is aware of the surrounding environment. To connect the external controllers (steering and pedals) to the CarMaker, an extension package from IPG called as Cockpit Package was used. To reproduce realistic behaviors of the surrounding traffic in the virtual environment, a co-simulation framework developed in our previous work was used \citep{Chada2021_1}. The co-simulation is performed by exchanging information about the traffic objects in real-time between traffic simulation software, SUMO and the IPG CarMaker. Matlab is used as the main interface tool for connecting both SUMO and the IPG CarMaker through TraCI Python and CarMaker4Simulink (CM4SL) APIs respectively.  Moreover, the vehicle model in IPG CarMaker was parameterized to replicate a Nissan Leaf BEV. The driver-in-the-loop (DiL) simulation runs on a 64-Bit Windows 10 PC equipped with Intel Core i9-10850K CPU, 3.9 GHz clock frequency and 64 GB RAM. The continuous eco-driving suggestions from pEDAS are displayed to the driver on a heads-up display (Figure~\ref{Fig:EDAS_suggestions}) and the driver controls the host vehicle while tracking these suggestions.
		\subsection{Participants}
	A total of 44 participants (Male=37, Female=7) possessing a valid EU driving license and normal or corrected-to-normal visual acuity volunteered to test the predictive eco-driving assistance system (pEDAS). Among these, 3 participants experienced motion sickness and therefore could not take part in the study. Hence, the survey feedback of N=41 participants was considered for the subjective analysis. While driving with pEDAS, 7 participants experienced technical issues during the test run, therefore their objective variables could not be compared. Hence, the sample size for the objective analysis has been reduced to N=34. The participants took a short socio-demographic survey in which the information regarding their age, gender, driving experience and driving style they possess (20.54\% cautious, 58.82\% normal or 20.54\% sporty) was gathered. Additionally, a feedback on the participant's familiarity with eco-driving concepts was obtained. The feedback revealed that around 2.5\% of the participants have never heard of eco-driving before, 42.5\% of the participants have heard about eco-driving but have never used it, 32.5\% are moderately familiar with similar eco-driving systems and have used it in few instances, 15\% are quite familiar with similar eco-driving systems and have used it occasionally and 7.5\% regularly use or have used a similar system while driving.
	\subsection{Experimental Procedure}
\begin{figure}
		\centering
		\includegraphics[width=0.8\linewidth]{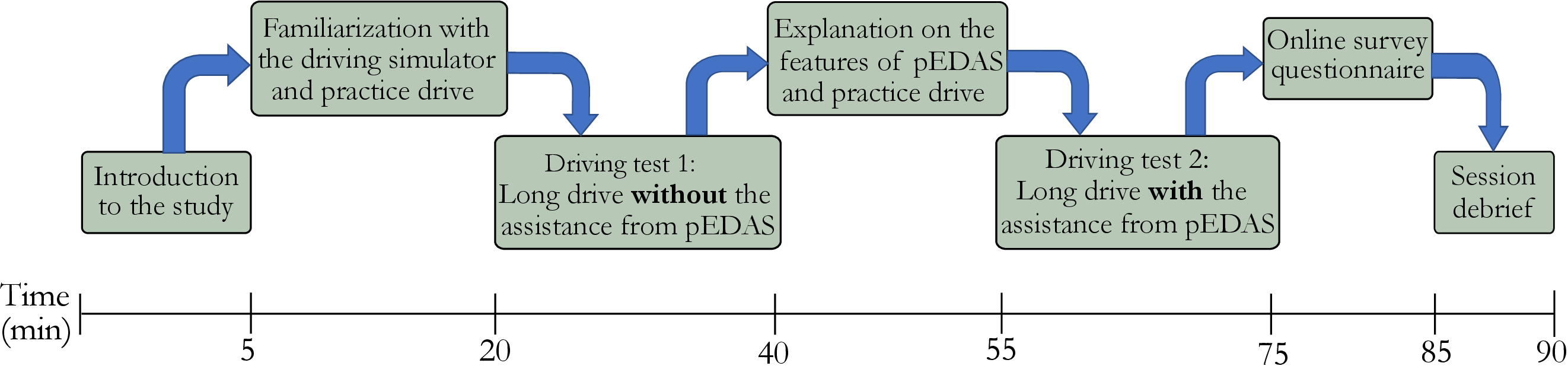}
		\caption{Experimental procedure}
		\label{fig:Experimental_procedure}
	\end{figure}

	The timeline of the experimental study is illustrated in Figure~\ref{fig:Experimental_procedure}. Upon arrival, participants were briefed about the present study and its objectives. It is important to mention here that pEDAS and its features were not yet introduced to the participants. In the first step, the drivers were presented with a short route and multiple trial drives to get familiarized with the driving simulator and its controls. In the next step, participants drove along a determined route of \unit[13.2]{km} consisting of urban and highway road segments in the presence of a realistic traffic, deterministic TL signals and stop signs. It is to be mentioned that the participants drove in this step with their own driving style and the assistance from pEDAS was not provided. To maintain a fair evaluation, participants were asked not to overtake a vehicle and also to maintain a minimum speed of \unit[80]{km/h} on the highway segment. After this drive, participants were introduced to pEDAS and its features as presented in Section~\ref{sec:Ecodriving_Feedback_System}. It is then followed by practice test drives where the participants got familiarized with pEDAS suggestions. After the practice drive, participants drove on the determined route, but now with the assistance from pEDAS. In the final step, participants answered an online survey questionnaire as presented in Appendix~\ref{appendix_A} (Tables \ref{appendix:tab2}, \ref{appendix:tab1} and \ref{appendix:tab3}).	   
	\subsection{Hypothesis}
	\label{sec:hypothesis}
	To predict the user acceptance of pEDAS in this work, following hypotheses are proposed based on TAM \citep{Davis1985} and TPB \citep{Ajzen1991}. TAM is formulated based on the Theory of Reasoned Action (TRA). The two important constructs of TAM are perceived usefulness (PU) and perceived ease of use (PEoU), where PU is defined as the degree to which an individual believes that using a specific system will enhance their job performance and PEoU is defined as the degree to which the individual expects that using the system to be free of effort. The TPB model is also based on TRA and serves as an extension to TAM. This model overcomes the limitation of TAM in dealing with behaviours over which people have incomplete volition control and improves the predictive ability \citep{Ajzen1991}. TPB proposes that behaviour intention and perceived behavioural control jointly determine the actual use behaviour rather than just attitude toward behaviour (ATT) which determines the actual use in TAM.
	\begin{myenv}
		\textit{\textbf{H1:} Perceived Ease of Use (PEoU) is a significant positive predictor of Perceived Usefulness (PU), implying that the individuals who believe that using pEDAS would be free of effort are more likely to believe that using pEDAS would enhance their driving performance.}
	\end{myenv}
	\begin{myenv}	
		\textit{\textbf{H2:} Perceived Usefulness (PU) has a positive significant effect on Attitude towards behaviour (ATT), implying that the individuals who believe that using pEDAS would enhance their driving performance are more likely to possess positive feelings towards pEDAS behavior.}
	\end{myenv}
	\begin{myenv}
		\textit{\textbf{H3:} Attitude toward behaviour (ATT) has a positive significant effect on Behavioural Intention (BI), implying that the individuals who possess positive feelings towards pEDAS behavior are more likely to intend to use pEDAS in daily life.}
	\end{myenv}	
	\begin{myenv}
		\textit{\textbf{H4:} Perceived Ease of Use (PEoU) has a positive significant effect on Attitude toward behaviour (ATT), implying that the individuals who believe that using pEDAS would be free of effort are more likely to possess positive feelings towards pEDAS behavior.}
	\end{myenv}	
	\begin{myenv}
		\textit{\textbf{H5:} Perceived Usefulness (PU) has a positive significant effect on Behavioural Intention (BI), implying that the individuals who believe that using pEDAS would enhance their driving performance are more likely to intend to use pEDAS in daily life.} 
	\end{myenv}	
	\begin{myenv}
		\textit{\textbf{H6:} Subjective Norms (SN) have a positive significant effect on Behavioural Intention (BI), implying that the individuals who believe that important people in their social network support using pEDAS are most likely to intend to use pEDAS in daily life.} 
	\end{myenv}
	\begin{myenv}		
		\textit{\textbf{H7:} Perceived Behavioral Control (PBC) has a positive significant effect on Behavioural Intention (BI), implying that the individuals who perceive ease in using pEDAS in the presence of necessary resources are most likely to intend to use pEDAS in daily life.} 	
	\end{myenv}	
	
	\section{Evaluation Results}
	To evaluate the driving performance and the user acceptance of the proposed predictive eco-driving assistance system (pEDAS), both objective and subjective evaluations were carried out in this work. The objective evaluation was conducted by defining certain key performance indicators (KPIs), specifically energy savings, reduction in speed limit violations and number of TL signals crossed at green. Moreover, driving behavior analysis was performed by comparing a participant's driving data before and after using pEDAS. The results of the objective evaluation are discussed in Section \ref{sec:objective_evaluation}. Furthermore, in Section \ref{sec:subjective_evaluation}, subjective evaluation was conducted to assess how the participants perceived pEDAS and to determine the factors that influence the acceptance of pEDAS amongst users.
	\subsection{Objective Evaluation}
	\label{sec:objective_evaluation}
	\subsubsection{Energy Savings}
\begin{figure*}
		\centering
		\begin{subfigure}[b]{0.475\textwidth}
			\centering
			\includegraphics[width=\textwidth]{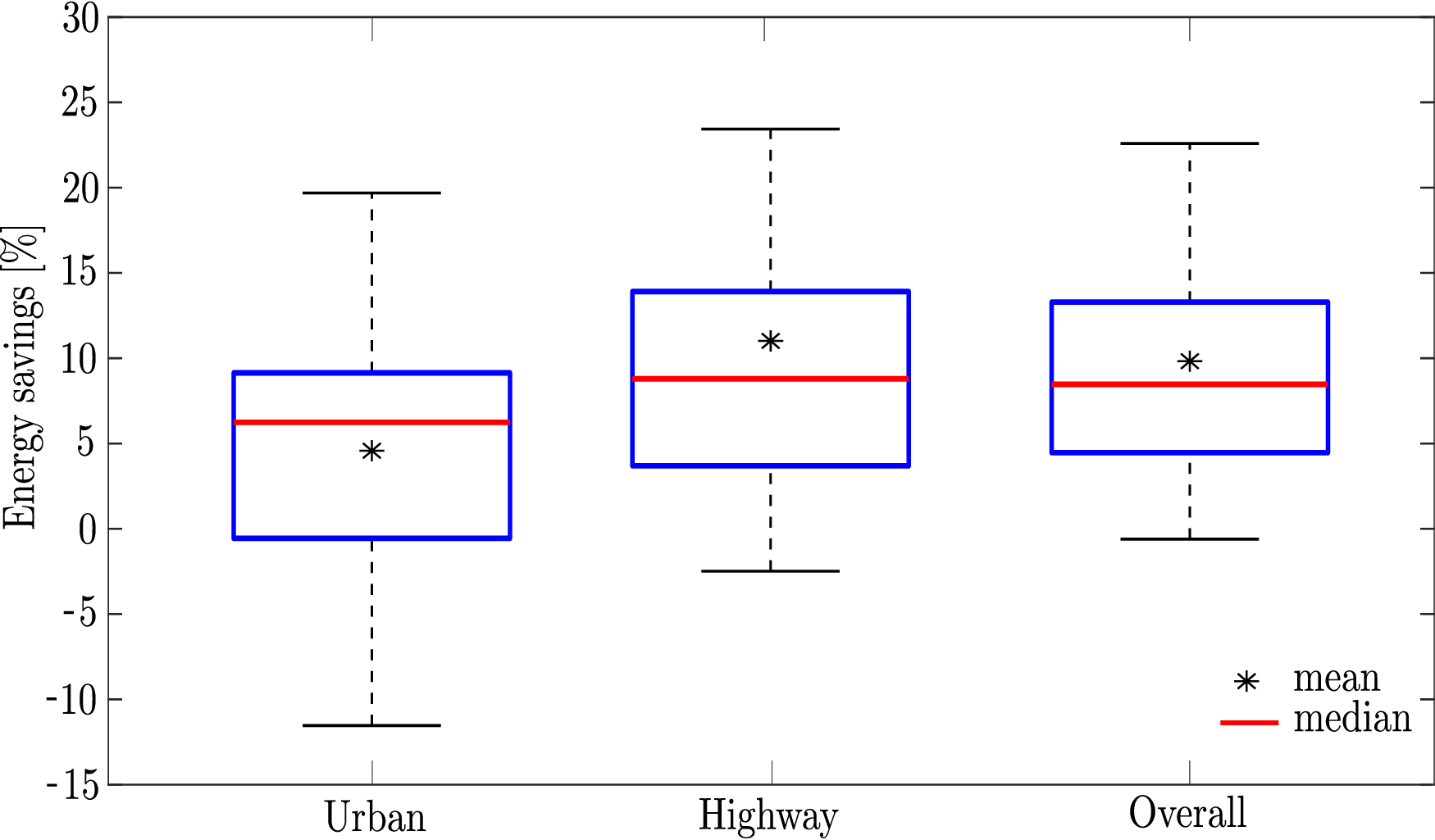}
			\caption[Network2]%
			{{\small Energy savings in the driven route}}    
			\label{fig:plot1}
		\end{subfigure}
		\hfill
		\begin{subfigure}[b]{0.475\textwidth}  
			\centering 
			\includegraphics[width=\textwidth]{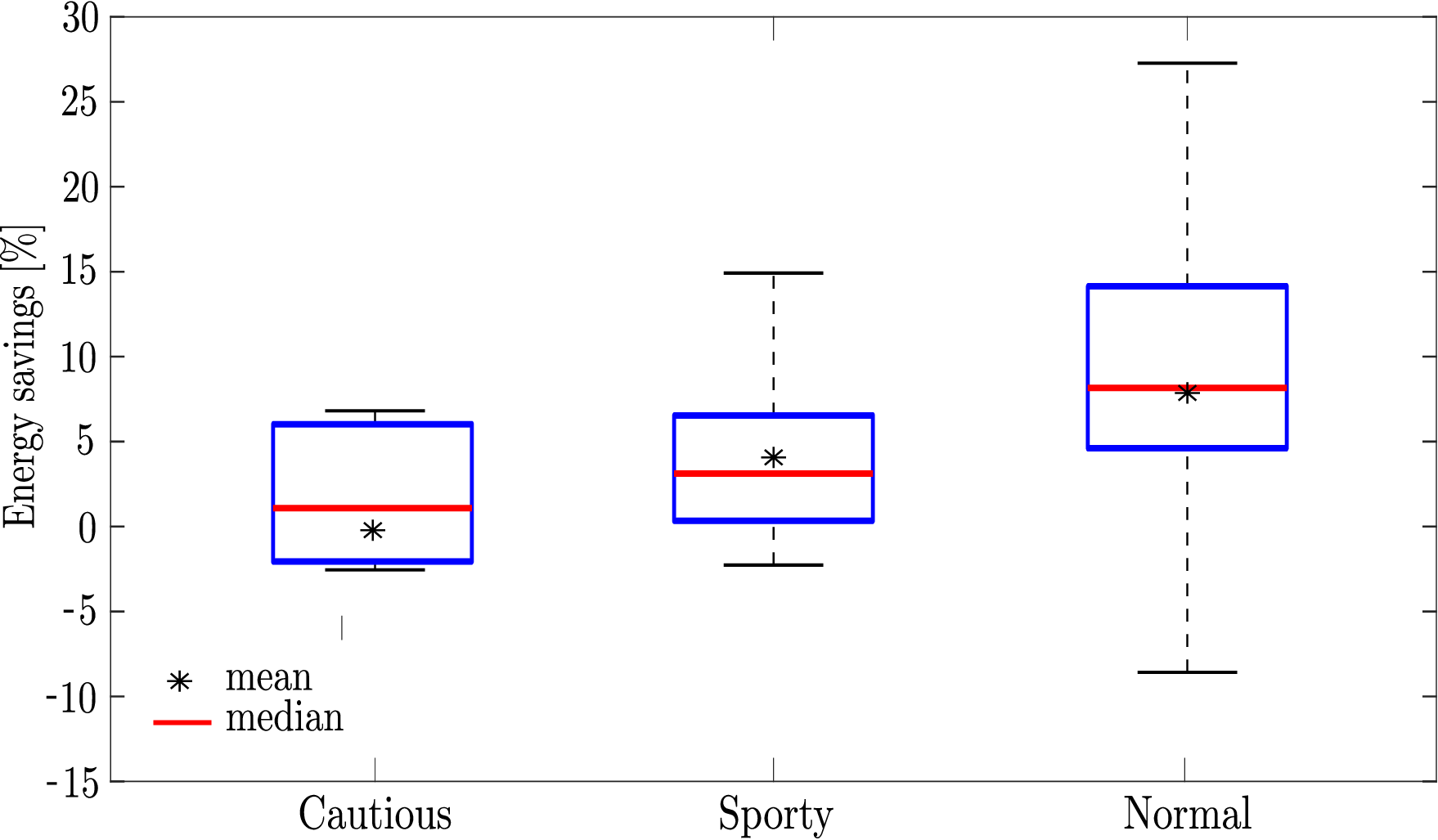}
			\caption[]%
			{{\small Energy savings according to driving styles: urban trip}}    
			\label{fig:plot2}
		\end{subfigure}
		\vskip\baselineskip
		\begin{subfigure}[b]{0.475\textwidth}   
			\centering 
			\includegraphics[width=\textwidth]{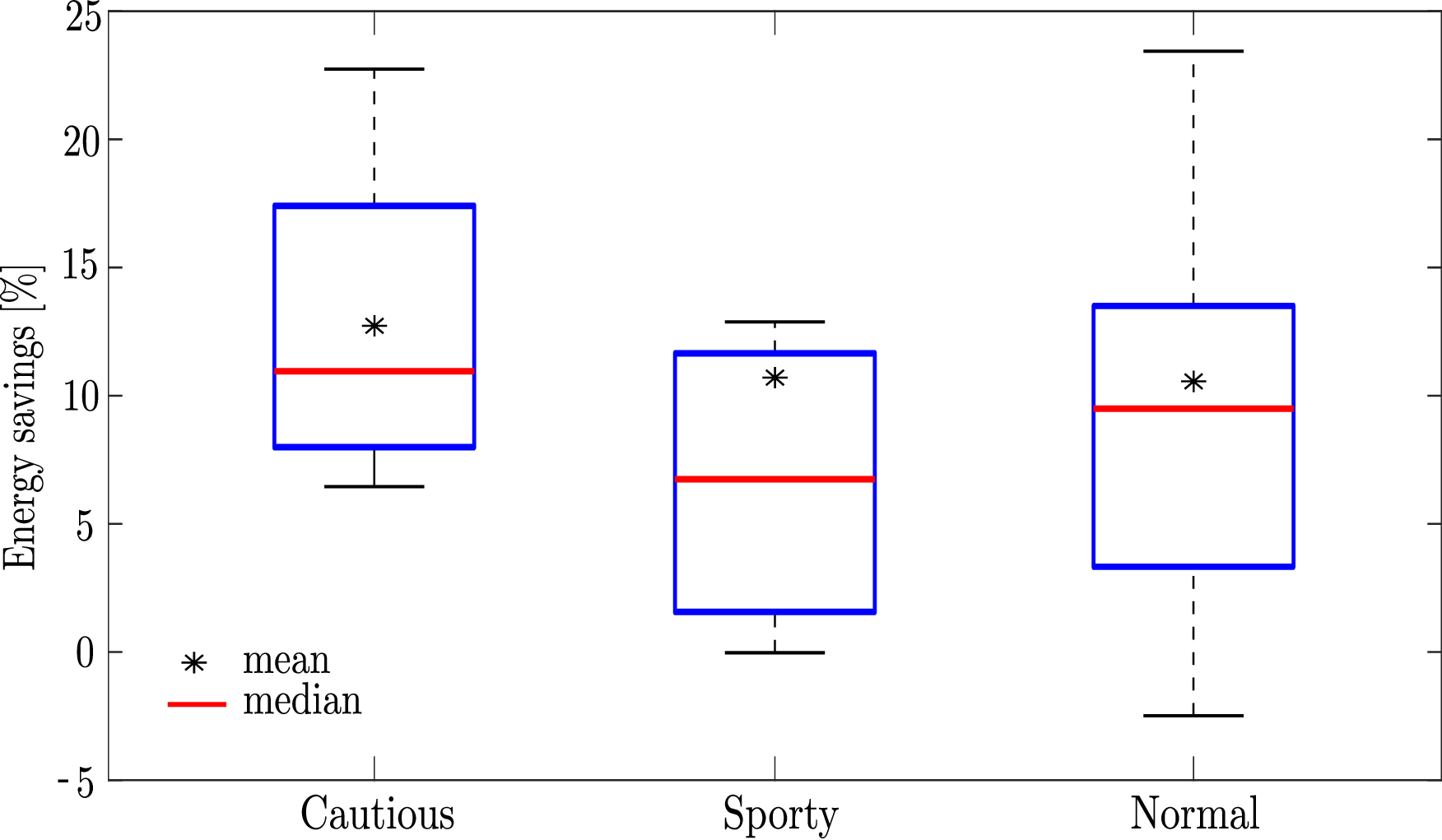}
			\caption[]%
			{{\small Energy savings according to driving styles: highway trip}}    
			\label{fig:plot3}
		\end{subfigure}
		\hfill
		\begin{subfigure}[b]{0.475\textwidth}   
			\centering 
			\includegraphics[width=\textwidth]{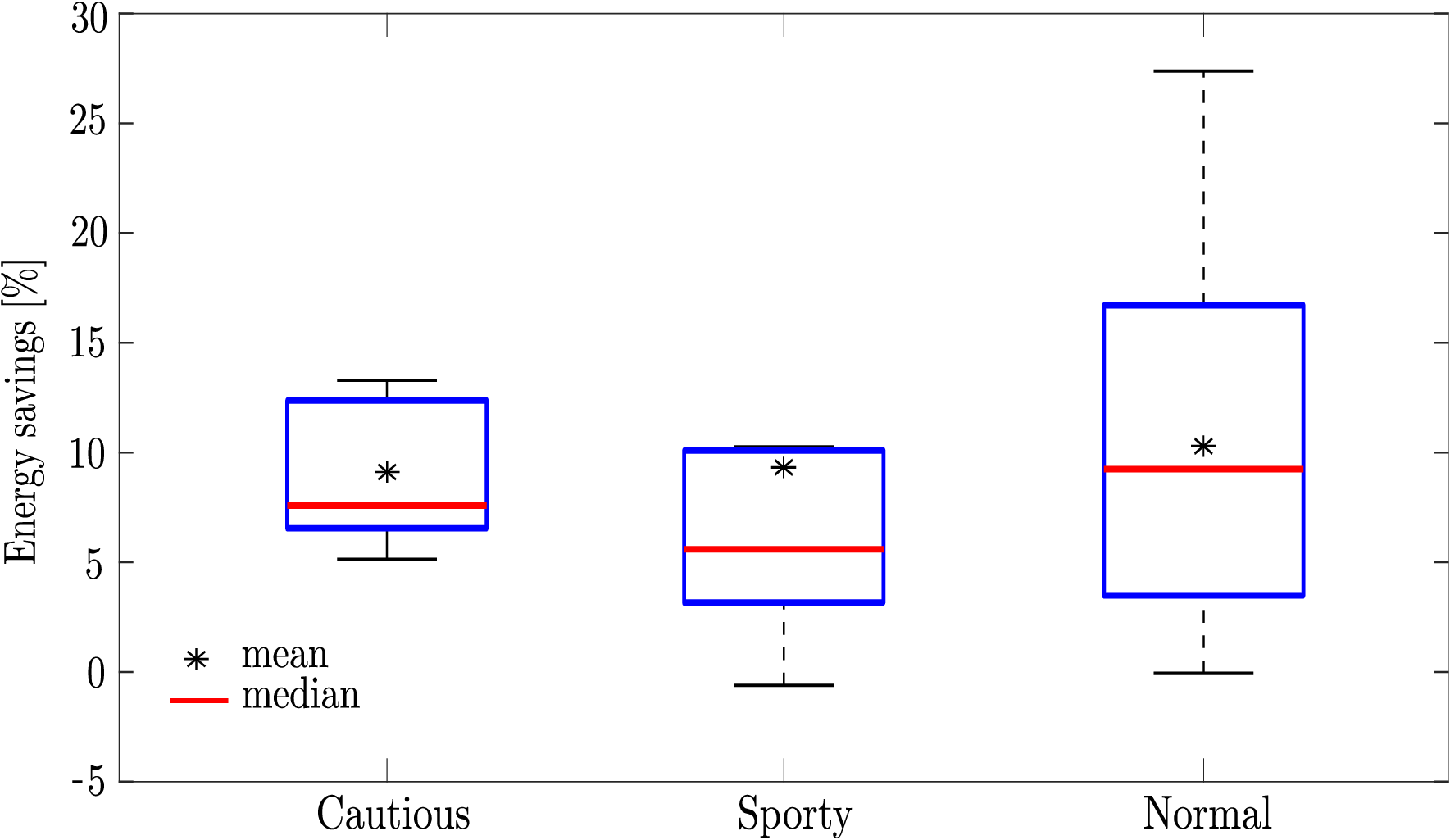}
			\caption[]%
			{{\small Energy savings according to driving styles: overall trip}}    
			\label{fig:plot4}
		\end{subfigure}
		\caption[ The average and standard deviation of critical parameters ]
		{\small Comparison of energy savings of participants} 
		\label{fig:mean and std of nets}
	\end{figure*}

	In Figure~\ref{fig:plot1}, the summary of the energy savings obtained for the urban, highway and overall trip is illustrated using a box plot. The participants driving with the continuous feedback from pEDAS have achieved average overall energy savings of 9.82\% (median 8.46\%) for the whole trip, an average savings of 4.6\% (median 6.24\%) in the urban segment and highest savings in highway segment with an average of 11\% (median 8.8\%) as compared to without the assistance from pEDAS. The energy savings in the urban segment are less compared to the highway segment due to the fact that the dense nature of the urban traffic has resulted in frequent acceleration and deceleration of the host vehicle that in turn affected the energy savings. The data also indicates that a few drivers could not save energy with the assistance from pEDAS. Further classifying the energy savings in the urban, highway and overall trips according to cautious, sporty and normal driving styles (as illustrated in Figure~\ref{fig:plot2}, \ref{fig:plot3} and \ref{fig:plot4} respectively), explained which type of drivers did not achieve significant savings. In urban segment (Figure~\ref{fig:plot2}), drivers with normal driving style achieved more energy savings with an average of 6.52\% (median 7.8\%), followed by drivers with sporty style with an average of 4\% (median 3.12\%). The cautious drivers were able to save comparatively less since their driving style was on par with pEDAS. One can observe from the highway trip in Figure~\ref{fig:plot3} that pEDAS in general has helped drivers from all categories of driving style to save energy, especially for drivers with normal driving style who achieved an average savings of 10.28\% (median 9.24\%). On the other hand, drivers with sporty and cautious styles were able to save comparatively less with an average savings of 9.25\% (median 5.59\%) and 9.08\% (median 7.58\%) respectively.
	\subsubsection{Road Speed Limit Violations} 
	The overall speed limit violations for the trip is reduced by almost half with an average reduction of 46.18\% (median 50.5\%) as a result of using pEDAS. The system helped the drivers to be aware of overspeeding and as a result reduced the overall mean speed which indirectly had an impact on reducing the energy consumption.
	\subsubsection{Crossing through Green Traffic Light Signals}  
\begin{figure}
		\centering
		\includegraphics[width=0.8\linewidth]{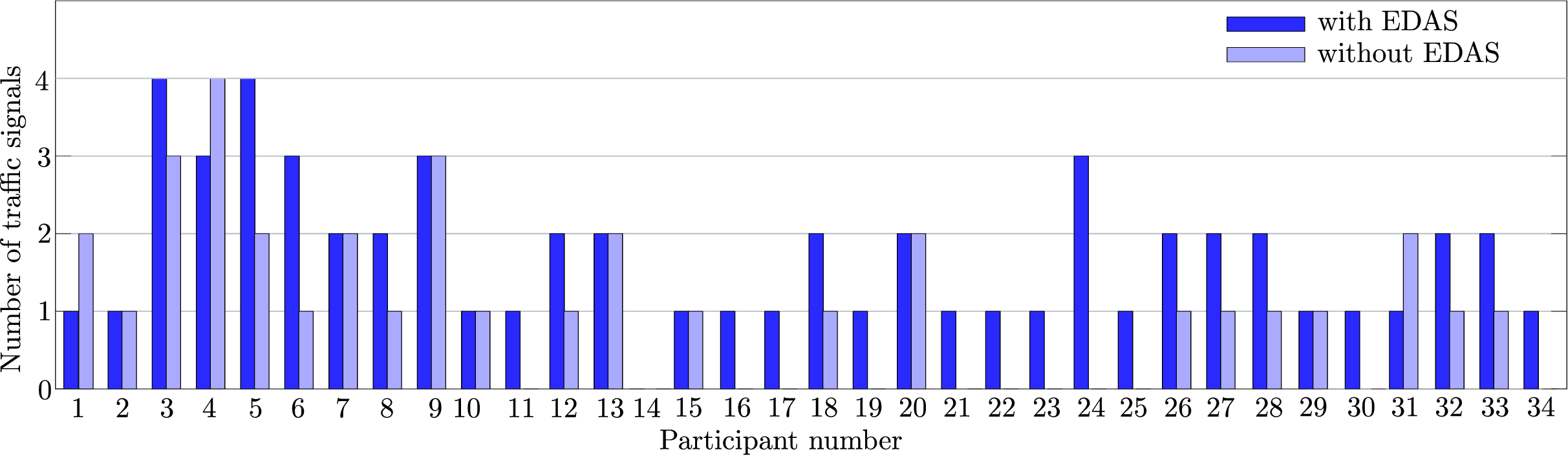}
		\caption{Comparison of traffic light signals crossed in green phase}
		\label{fig:plot_crossing_green_TLs}
	\end{figure}

	Figure~\ref{fig:plot_crossing_green_TLs} compares the number of traffic light (TL) signals crossed at a green phase for all the participants with and without the assistance from pEDAS. With the green-wave algorithm introduced in Section \ref{sec:green_wave_velcoity_calculation}, the chances of crossing the TL signals at green phase in general have increased. Overall, stopping at red TL signals has been decreased by 60\% on the average when using pEDAS. The empty entries in Figure~\ref{fig:plot_crossing_green_TLs} indicate that the drivers have stopped at all TL signals. In some cases drivers have stopped at more TL signals with pEDAS as well. It is important to note here that this is not due to the ineffectiveness of the green-wave algorithm but rather due to the stochastic nature of the traffic, which has resulted in unfavorable scenarios. For example, encountering a preceding vehicle will deactivate the green-wave feature and can increase the chances of stopping at a red phase. It is evident from the results that under right circumstances, the green-wave algorithm helps to maximize the chances of crossing at a green phase.
\subsubsection{Driving Behavior Analysis} 
\begin{figure}[t]
	\centering
	\includegraphics[width=1.0\linewidth]{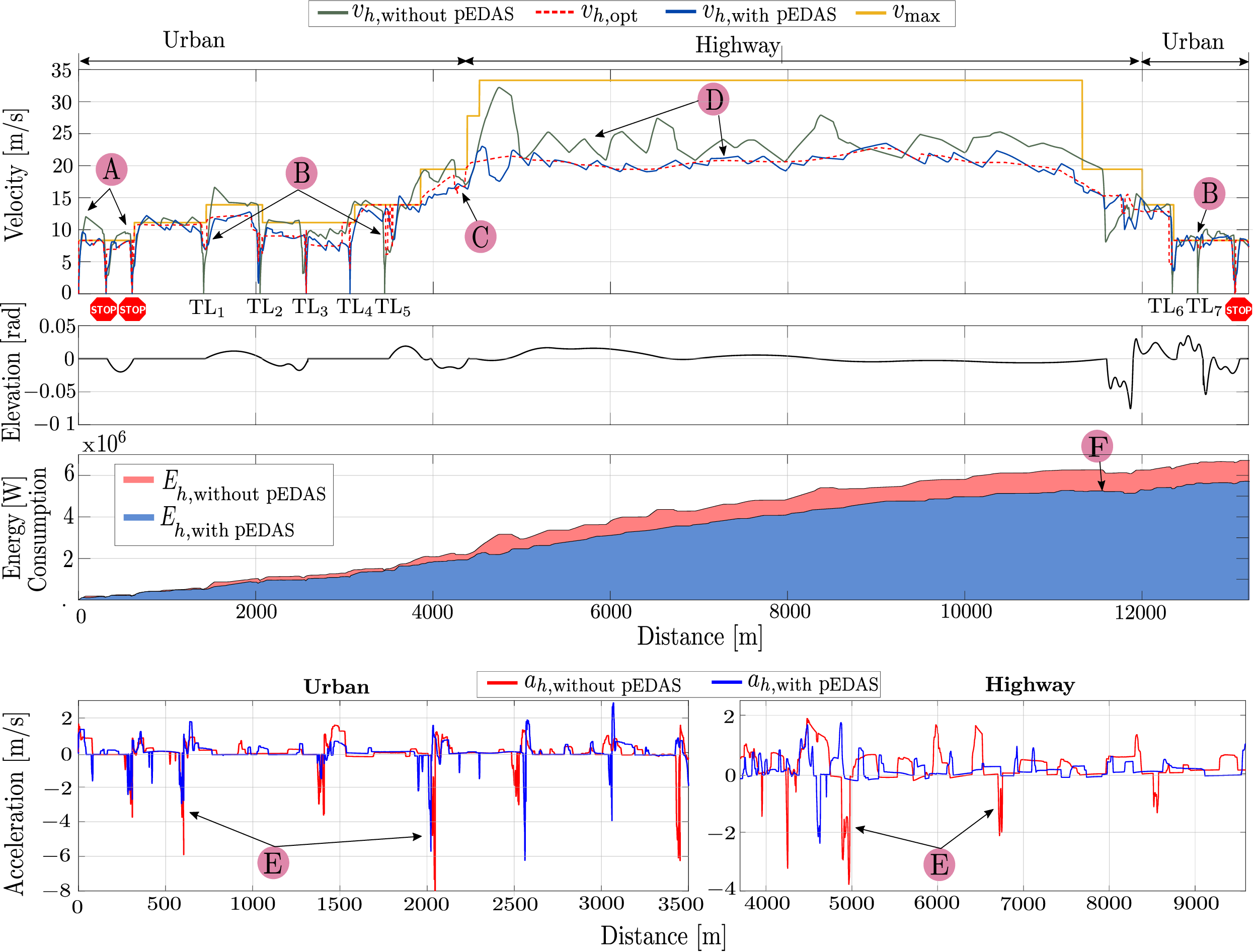}
	\caption{Driving behavior comparison of participant $6$ before and after using pEDAS through velocity, energy consumption and acceleration/deceleration profiles}
	\label{fig:TRC_objective_velocity}
\end{figure}

	In Figure~\ref{fig:TRC_objective_velocity}, the driving behavior of one of the participants before and after using pEDAS is compared through the velocity, acceleration and energy consumption graphs. It can be observed that pEDAS has assisted the driver to (A) not exceed the speed limits $v_\text{max}$ quite often, (B) cross through three signalized intersections $\text{TL}_{1}$, $\text{TL}_{5}$ and $\text{TL}_{7}$ at a green phase by tracking a green-wave optimal speed, (C) travel at safe and lowered speeds at curved roads, (D) reduce the overall average driving speeds, (E) lower unnecessary accelerations and decelerations and (F) thereby reduce the BEV energy consumption as compared to without the assist from pEDAS. The participant was able to achieve overall energy savings up to 15\%, in which the urban part constitutes to 7\% and the highway part constitutes to 18.5\% of energy savings. Moreover, as a consequence of reduced acceleration and decelerations while tracking the suggestions from pEDAS, the overall jerk has been reduced by 12\%. It is evident from the aforementioned observations that the participant has benefited from abiding to the pEDAS recommendations and improved the driving style in energy-efficient manner. 
\begin{figure}[t]
\centering
\includegraphics[width=1.0\linewidth]{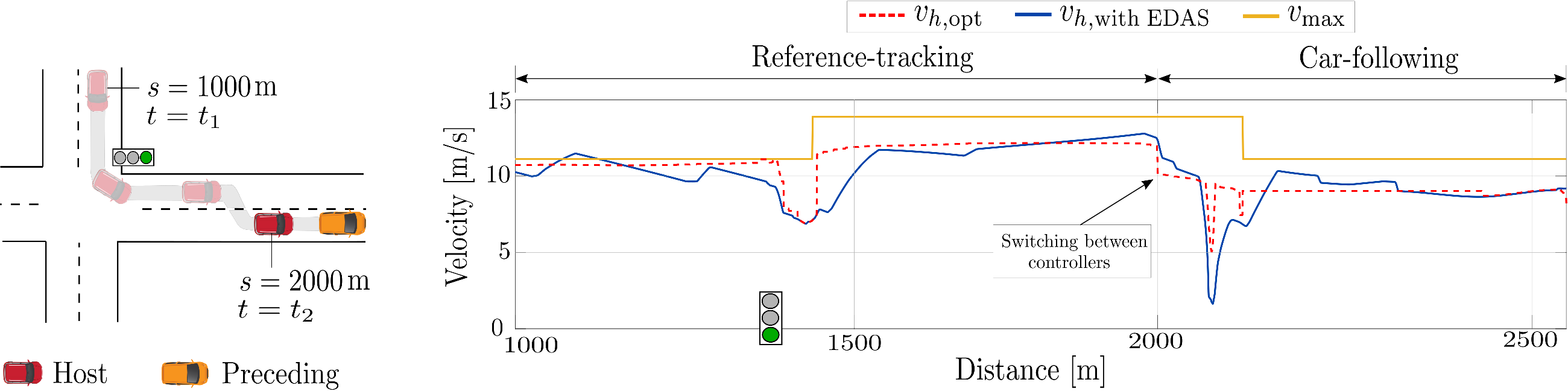}
\caption{Scenario obtained from participant 6 to demonstrate the switching between reference-tracking and car-following controllers }
\label{fig:switching_controllers_example}
\end{figure}	

To demonstrate the switching performance between reference-tracking and car-following controller, a scenario encountered while driving with pEDAS is shown in Figure~\ref{fig:switching_controllers_example}. The left illustration depicts the scenario in which at time $t = t_{1}$ and distance $s=\unit[1000]{m}$, the host car equipped with pEDAS is in free-way driving, and tracks a green-wave optimal speed using the reference-tracking MPC to cross the TL signal at a green-phase. It can be noticed from the right illustration in Figure~\ref{fig:switching_controllers_example} that after crossing the TL signal at $s=\unit[1400]{m}$, the $v_{h,\text{opt}}$ recommended a speed reduction at a turn. Later at time $t = t_{2}$ and distance $s=\unit[2000]{m}$, as the host car merges to the adjacent lane, the Eco-MPC controller switches to the car-following controller due to the presence of a preceding vehicle ahead. The $v_{h,\text{opt}}$ recommends a new optimal speed to track the preceding vehicle. At the time of switching, the transitional behavior between the reference-tracking and car-following controllers can be observed to be smooth and stable. Moreover, further observations made in the driver behavior analysis for other participants before and after using pEDAS are summarized and discussed in Section~\ref{sec:discussion_summary_results}.
%

	\subsection{Subjective Evaluation}
	\label{sec:subjective_evaluation}
	To assess the user opinions and feelings towards the acceptance of pEDAS, a subjective evaluation in this work is carried out using the technology acceptance model (TAM) and theory of planned behavior (TPB). The participants responded to a survey questionnaire that included the items of the constructs perceived usefulness (PU), perceived ease of use (PEoU), attitude towards behavior (ATT), subjective norms (SN), perceived behavioral control (PBC) and behavioral intention (BI) as presented in Appendix~\ref{appendix_A} (Table~\ref{appendix:tab1} \& \ref{appendix:tab2}). The questionnaire items were measured using a 5-point Likert scale (1=strongly disagree, 3=neutral and 5=strongly agree). In this work, structural equation modeling (SEM) \citep{Maruyama1998, Ullman2012} was used to evaluate the correlations between latent constructs and verify the influence of several determinants on the behavioral intention to use pEDAS. The SEM approach comprises of two steps, firstly, a measurement model analysis and structural model analysis \citep{Maruyama1998}. To perform the statistical analysis in this work, IBM SPSS 28 was used. 
	\subsubsection{Measurement Model Analysis}
	The measurement model deals with the relationship between the latent constructs and its indicators \citep{Ullman2012}. To validate the goodness-of-fit of the measurement model, a confirmatory factor analysis must be conducted. In this analysis, the reliability and validity of the items in terms of indicator reliability, construct reliability, convergent validity and discriminant validity are tested \citep{Gunthner2021}. Indicator reliability represents the proportion of the variance in an item that is explained by the latent construct \citep{Hair2014}. It can be explained from the standardized factor loading $\lambda$. According to \citet{Hair2014}, the items with the factor loading above the desired threshold of 0.7 must be maintained. To estimate the internal consistency of the items, construct/composite reliability (CR) and Cronbach's alpha (C$\alpha$) are widely used methods in the literature \citep{Hair2014}. CR is considered as an appropriate measure as compared to C$\alpha$, as the former assumes the varying factor loading in contrast to the latter \citep{Hair2014}. The literature recommends the CR and the C$\alpha$ values to be greater than 0.6 and 0.7 respectively. Convergent validity is defined as the extent to which an item correlates positively with other items of the same latent variable. To test a convergent validity, average variance extracted (AVE) is used. An AVE value of greater than 0.5 is a must for the latent variable to be able to explain more than 50\% of the variance of its items \citep{Hair2014, Gunthner2021}. Discriminant validity is defined as the extent to which a latent variable is unique and less correlated with other latent variables. To validate the discriminant validity, a criteria suggested in \cite{Fornell1981} is used, which states that the square root of AVE of each latent variable must be larger than its correlation with other latent variables.
	\begin{table}[width=1\linewidth,pos=t]
		\caption{Descriptive statistics along with standardized factor loading, cronbach's alpha, composite reliability and average variance extracted}\label{tbl3}
		\centering
		\begin{adjustbox}{width=\columnwidth,center}
			\begin{threeparttable}
				\begin{tabular}{>{\centering\arraybackslash}p{0.4\textwidth}>{\centering\arraybackslash}p{0.1\textwidth}>{\centering}p{0.1\textwidth}>{\centering}p{0.2\textwidth}>{\centering}p{0.1\textwidth}>{\centering}p{0.1\textwidth}>{\centering\arraybackslash}p{0.1\textwidth}}
					\toprule
					Constructs                       & Mean & SD     & $\lambda$  & $C\alpha$ & CR & AVE \\
					\midrule
					\textbf{Perceived Usefulness(PU)}         &           &            &                 & 0.799            & 0.879                 & 0.646                      \\
					PU1                                   & 4         & 0.847      & 0.837           &                  &                       &                            \\
					PU2                                   & 4.35      & 0.735      & 0.792           &                  &                       &                            \\
					PU3                                   & 3.625     & 1.21       & 0.747           &                  &                       &                            \\
					PU4                                   & 4.05      & 0.959      & 0.837           &                  &                       &                            \\
					\textbf{Perceived Ease of Use (PEoU)}        &           &            &                 & 0.596            & 0.772                 & 0.537                      \\
					PEoU1                                 & 4.125     & 0.882      & 0.724           &                  &                       &                            \\
					PEoU2                                 & 3.75      & 0.839      & 0.585           &                  &                       &                            \\
					PEoU3\tnote{\ding{61}}                                 & 3.65      & 1.026      & -0.193\tnote{*}          &                  &                       &                            \\
					PEoU4                                 & 4.35      & 0.833      & 0.863           &                  &                       &                            \\
					\textbf{Attitude of Behavior (ATT)}         &           &            &                 & 0.828                 & 0.873                 & 0.634                      \\
					ATT1                                  & 4.25      & 0.808	   & 0.464\tnote{*}           &                  &                       &                            \\
					ATT2                                  & 4.225     & 0.861      & -0.034\tnote{*}          &                  &                       &                            \\
					ATT3                                  & 4.05      & 0.959      & 0.249\tnote{*}           &                  &                       &                            \\
					ATT4                                  & 4.2       & 0.648      & 0.687\tnote{*}           &                  &                       &                            \\
					ATT5                                  & 3.85      & 0.975      & 0.579\tnote{*}           &                  &                       &                            \\
					ATT6                                  & 3.8       & 0.822      & 0.817           &                  &                       &                            \\
					ATT7                                  & 3.575     & 0.957      & 0.820           &                  &                       &                            \\
					ATT8                                  & 3.725     & 1.085      & 0.802           &                  &                       &                            \\
					ATT9                                  & 4.375     & 0.627      & 0.743           &                  &                       &                            \\
					\textbf{Perceived Behavioral Control (PBC)} &           &            &                 & 0.837            & 0.904                 & 0.759                      \\
					PBC1                                  & 4.075     & 0.916      & 0.869           &                  &                       &                            \\
					PBC2                                  & 4.208     & 0.779      & 0.858           &                  &                       &                            \\
					PBC3                                  & 4.4       & 0.744      & 0.888           &                  &                       &                            \\
					\textbf{Subjective Norms (SN)}             &           &            &                 & 0.917            & 0.949                 & 0.862                      \\
					SN1                                   & 3.675     & 1.095      & 0.92            &                  &                       &                            \\
					SN2                                   & 3.675     & 0.971      & 0.936           &                  &                       &                            \\
					SN3                                   & 3.525     & 0.933      & 0.93            &                  &                       &                            \\
					\textbf{Behavioral Intention (BI)}         &           &            &                 & 0.792            & 0.907                 & 0.829                      \\
					BI1                                   & 3.525     & 1.03       & 0.911           &                  &                       &                            \\
					BI2                                   & 4.175     & 0.930      & 0.911           &                  &                       &                           
					\\ 			\bottomrule                                             
				\end{tabular} 
				\begin{tablenotes}
					\item[]{$\lambda$ = Factor loading, $C\alpha$ = Cronbach's Alpha, CR = Composite/Construct Reliability, AVE = Average Variance Extracted,}
					\item[*]{omitted from the analysis}
					\item[\ding{61}]{reverse scaled item}	
				\end{tablenotes}
			\end{threeparttable}
		\end{adjustbox}
	\end{table}
	\begin{table}[width=1\linewidth,pos=t]
		\caption{Inter-construct correlation matrix of the TAM and TPB constructs}\label{tbl4}
		\centering
		\begin{adjustbox}{width=\columnwidth,center}
			\begin{threeparttable}
				\begin{tabular}{>{\centering\arraybackslash}p{0.15\textwidth}>{\centering\arraybackslash}p{0.15\textwidth}>{\centering}p{0.15\textwidth}>{\centering}p{0.15\textwidth}>{\centering}p{0.15\textwidth}>{\centering}p{0.15\textwidth}>{\centering\arraybackslash}p{0.15\textwidth}}
					\toprule
					Factors   & PU & PEoU     & ATT  & PBC  & SN    & BI    \\
					\midrule
					PU       & $\mathbf{0.803\tnote{\ding{61}}}$            &     			&       &       &       &       \\
					PEoU     &  0.450\tnote{**}  &  $\mathbf{0.641\tnote{\ding{61}}}$ 		&       &       &       &       \\
					ATT      &  0.772\tnote{**}  & .328\tnote{*} &  $\mathbf{0.796\tnote{\ding{61}}}$ &       &       &       \\
					PBC       &  .376\tnote{*}    &   .595\tnote{**}     &  0.106     &  $\mathbf{0.871\tnote{\ding{61}}}$ &       &       \\
					SN        &  .492\tnote{**}   &  -0.041     &   .460\tnote{**}     &  -0.059    &  $\mathbf{0.928\tnote{\ding{61}}}$ &       \\
					BI        &  .622\tnote{**}   &   .312\tnote{*}    &    .395\tnote{*}   &  .495\tnote{**}         &  .352\tnote{*}     &  $\mathbf{0.91\tnote{\ding{61}}}$ \\ 
					\bottomrule
				\end{tabular}
				\begin{tablenotes}
					\item[\ding{61}]{square root of AVE}
					\item[**]{Correlation is significant at the 0.01 level (2-tailed).}
					\item[*]{Correlation is significant at the 0.05 level (2-tailed).}
				\end{tablenotes}
			\end{threeparttable}
		\end{adjustbox}
	\end{table}
	The descriptive statistics (N=41) with mean M and standard deviation SD for each item along with the standardized factor loading $\lambda$, cronbach's alpha C$\alpha$, composite reliability CR and average variance extracted AVE are presented in Table~\ref{tbl3}. Furthermore, the bi-variate correlations between the latent constructs are summarized in the Table~\ref{tbl4}. The $\lambda$ for the items PEOU2-PEOU3 and ATT1-ATT5 is found to be less than 0.7 (Table~\ref{tbl3}), however more evidence is necessary to delete the aforementioned items. The authors have noticed that the items of the construct ATT were highly correlating with the items of PU, which is not intended. Thus, omitting the items ATT1-ATT5 from the analysis resulted in satisfying the discriminant validity criteria for the inter-correlation construct ATT and other constructs as well (Table~\ref{tbl4}).To maintain the consistency with the scale, the reverse scaled item PEoU3 was reversed. Furthermore, deleting the negative loaded item PEoU3 (Table~\ref{tbl3}) has improved the AVE of the construct PEoU (0.537). It can be observed from Table~\ref{tbl3}, that the CR and AVE are higher than the desired values of 0.6 and 0.5 respectively. The C$\alpha$ for all the other constructs is found to be greater than 0.7 except for PEoU (0.596), which shows a moderate reliability as the value lies between 0.5-0.7 \citep{Hinton2004}. Deleting the item PEOU2 did not improve the C$\alpha$ much, therefore the authors have retained this item. 
	\subsubsection{Structural Model Analysis}
	\begin{table}[width=1\linewidth,pos=t]
		\caption{Hierarchical regression analysis on TAM and TPB models}\label{tbl5}
		\centering
		\begin{adjustbox}{width=\columnwidth,center}
			\begin{threeparttable}
				\begin{tabular}{p{0.3\textwidth}>{\centering\arraybackslash}p{0.12\textwidth}>{\centering}p{0.12\textwidth}>{\centering}p{0.12\textwidth}>{\centering}p{0.12\textwidth}>{\centering}p{0.12\textwidth}>{\centering}p{0.12\textwidth}>{\centering\arraybackslash}p{0.12\textwidth}}
					\toprule
					Models                           &$R^2$  & Adj. $R^2$ & $\Delta R^2$ & B     & SE B  & $\beta$  & 95\% $CI$          \\
					\midrule
					1. PEoU $\rightarrow$ PU         &  &                          &            &                      &                      &                      &                      \\
					Step 1: Dependent Variable - PU   &0.161 & 0.139                    & 0.161      &                      &                      &                      &                      \\
					Independent Variable - PEoU       &  &                          &            &0.478                & 0.177                & 0.401\tnote{*}                & 0.217-1.031          \\
					&   &                       &            &                      &                      &                      &                      \\
					2. PEoU + PU $\rightarrow$ ATT     & &                          &            &                      &                      &                      &                      \\
					Step 1: Dependent Variable - ATT    &0.056 & 0.031                    & 0.056      &                      &                      &                      &                      \\
					Independent Variable - PEoU         & &                          &            & 0.270                & 0.180                & 0.237                 & 0.023-0.848          \\
					&                   &       &            &                      &                      &                      &                      \\
					Step 2: Dependent Variable - ATT    &0.602 & 0.580                    & 0.546      &                      &                      &                      &                      \\
					Independent Variable - PEoU        &  &                          &            & -0.099               & 0.129                & -0.087               & -0.361-0.162        \\
					Independent Variable - PU          &  &                          &            & 0.773                & 0.109                & 0.807\tnote{**}                 & 0.553-0.993          \\
					&                       &   &            &                      &                      &                      &                      \\
					3. ATT + PU $\rightarrow$ BI   &    &                          &            &                      &                      &                      &                      \\
					Step 1: Dependent Variable - BI   &0.156  & 0.133                    & 0.156      &                      &                      &                      &                      \\
					Independent Variable - ATT         & &                          &            & 0.488                & 0.184                & 0.395\tnote{*}                 & 0.115-0.862          \\
					&          &                &            &                      &                      &                      &                      \\
					Step 2: Dependent Variable - BI    &0.405 & 0.373                    & 0.249      &                      &                      &                      &                      \\
					Independent Variable - ATT         & &                          &            & -0.262               & 0.247                & -0.212               & -0.762-0.238         \\
					Independent Variable - PU          &  &                          &            & 0.931                & 0.236                & 0.785\tnote{**}                 & 0.452-1.410          \\
					&         &                 &            &                      &                      &                      &                      \\
					4. ATT + PBC + SN $\rightarrow$ BI & &                                                        &                      &                      &                      &                      &                      \\
					Step 1: Dependent Variable - BI    & \multicolumn{1}{c}{\multirow{2}{*}{same as model 3 step 1}} & & \multicolumn{1}{c}{} & \multicolumn{1}{c}{} & \multicolumn{1}{c}{} & \multicolumn{1}{c}{} & \multicolumn{1}{c}{} \\
					Independent Variable - ATT          & \multicolumn{1}{c}{}                                   &                   &   &                      &                      &                      &                      \\
					&                    &                                    &                      &                      &                      &                      &                      \\
					Step 2: Dependent Variable - BI     &0.193 & 0.149                    & 0.037      &                      &                      &                      &                      \\
					Independent Variable - ATT        &  &                          &            & 0.365                & 0.206                & 0.295                 & -0.052 - 0.782        \\
					Independent Variable - SN        &  &                          &            & 0.208                & 0.160                & 0.216                 & -0.116 - 0.533        \\
					&                          &    &        &                      &                      &                      &                      \\
					Step 3: Dependent Variable - BI    &0.425 & 0.377                    & 0.233      &                      &                      &                      &                      \\
					Independent Variable - ATT       &   &                          &            & 0.263                & 0.178                & 0.213                & -0.098 - 0.624       \\
					Independent Variable - SN      &    &                          &            & 0.273                & 0.138                & 0.283              &   -0.007 - 0.553      \\
					Independent Variable - PBC       &    &                          &            & 0.617                & 0.162                & 0.489\tnote{**}               &  0.289 - 0.944      \\ 			\bottomrule                                             
				\end{tabular} 
				\begin{tablenotes}
					\item[**]{Correlation is significant at the 0.001 level (2-tailed).}
					\item[*]{Correlation is significant at the 0.05 level (2-tailed).}
				\end{tablenotes}
			\end{threeparttable}
		\end{adjustbox}
	\end{table}
	To test the hypotheses H1-H7 discussed in Section~\ref{sec:hypothesis} and explain the variance contributed by the latent constructs on the behavioral intention or the actual use of the system, a hierarchical regression analysis (HRA) is performed in this study. Four models that use perceived usefulness (PU), attitude towards behavior (ATT) and behavioral intention (BI) as dependent variables and their corresponding results are summarized in Table~\ref{tbl5}. Performing the linear regression analysis on model 1 revealed that the construct perceived ease of use (PEoU) has a positive significant effect on PU with unstandardized coefficients B$ =0.478$, SE B$=0.177$, and standardized coefficient $\beta=0.401$. PEoU was able to explain 16.1\% of variance in PU (Adj$R^2 = 0.139$, $\Delta R^2=0.161$). Even though the proportion of explained variance in PU is small, it is statistically significant with $p < 0.05$, thus validating hypothesis 1.
	
	The hierarchical linear regression is performed on model 2, in which PEoU is inserted in to the model in step 1. Subsequently PU is added in to the model in the second step. In the first step, PEoU explained only 5.6\% of the variance in ATT (Adj$R^2 = 0.031$, $\Delta R^2=0.056$). In step 2, PU increased the variance by a significant 54.6\% ($p<0.01$), increasing the adjusted $R^2$ to 0.58 and $\Delta R^2$ to 0.546. Therefore, PU is found to have a positive significant effect on ATT with unstandardized coefficients B$ =0.773$, SE B$=0.109$ and standardized coefficient $\beta=0.807$, thus supporting hypothesis 2. However, the addition of PU in turn reduced the effect of PEoU on ATT ($\beta=-0.087$) and it was not found to be significant ($p>0.05$), thus hypothesis 4 is not supported. This shows that PU is a significant predictor of ATT above and beyond PEoU. 
	
	In the first step in model 3, ATT entered and was able to explain only 15.6\% of the variance in BI (Adj$R^2 = 0.133$, $\Delta R^2=0.156$). Subsequently when PU entered the model in step 2, the variance increased to 40.5\% (Adj$R^2 = 0.373$, $\Delta R^2=0.249$, $p<0.001$). Thus, PU was found to be a significant predictor of BI ($p < 0.001$) with unstandardized coefficients B$ =0.931$, SE B$=0.236$ and standardized coefficient $\beta=0.785$, thus supporting hypothesis 5. The addition of PU in turn reduced the effect of ATT on BI ($\beta=-0.212$, $p>0.05$), thus hypothesis 3 is not supported. This implies that PU alone is a significant predictor of BI above and beyond ATT.
	
	In the model 4, ATT entered the model in the step 1, subsequently subjective norms (SN) in the step 2 and perceived behavioral control (PBC) in the step 3. Adding the SN in step 2 in turn reduced the contribution of ATT ($\beta$ = 0.295) rendering it insignificant. Although, after adding SN to the model the variance increased to 19.3\% (Adj$R^2 = 0.149$, $\Delta R^2=0.037$, $p>0.05$), the change in the $R^2$ was found to be very small ($\Delta R^2$=0.037) and was not statistically significant. Thus, it can be concluded that SN is not a significant contributor in predicting the BI, thus hypothesis H6 is not supported. When PBC entered the model in step 3, the variance increased to 42.5\% (Adj$R^2 = 0.377$, $\Delta R^2=0.233$) and it was significant ($p<0.05$). This implies that PBC alone is a significant predictor of behavioural intention with B $=0.617$, SE B$=0.162$ and $\beta=0.489$  above and beyond SN and ATT. 
	
	\subsection{Discussion and Summary of the Results}
	\label{sec:discussion_summary_results}
The objective analysis revealed that the drivers were able to achieve an average energy savings of 9.82\% for the overall trip using pEDAS. This result is in line with the previous findings from the literature on eco-driving training and support systems \citep{Vagg2013,Ruben2021,Sullman2015}. Additionally, with the green-wave optimized speed, the chances of crossing the TL signals at green phase increased. It is expected that the energy savings could increase over time if the drivers implement the learned eco-driving techniques in daily life \citep{Sullman2015}. Furthermore, it was noticed from the driver behavior analysis, that the drivers in general have reduced mean driving speeds in the highway segment based on the recommendations from pEDAS, specifically by following an optimal speed calculated using DP (Figure~\ref{fig:DP_profile}), which resulted in reduced mean energy consumption of up to 11\%. Although the generated DP profile favored more energy savings, a few drivers complained about the low mean speeds on the highway segment. To address this in the future, generating multiple offline DP speed profiles by analyzing the driver's requirement on trip-time with a trade-off on energy savings can be considered. Furthermore, the data from the analysis showed that pEDAS has the potential to reduce mean energy consumptions in urban segments as well. Here the drivers with cautious driving style had less benefits or on par to the baseline due to their lower mean speeds. A point to be noted here is that the parameters of the optimization algorithm and the HMI are kept constant across all the participants. In case of a cautious driver, in some situations this resulted in modifying their driving behaviour which was not suiting them. Hence, learning the parameters such as time headway, prediction horizon, desired inter-vehicle distance from different driver types is necessary. This calls for a personalization of pEDAS that adapts the driver behavior and helps to achieve better energy savings. The importance of the adaptation of ADAS and HMI to the driver's preferences and driving style is discussed in \citet{Hasenjager2020}. It is worth to look at the extreme cases in the urban segment where drivers were not able to save or consumed more energy with assistance from pEDAS. In one such case for a driver with normal driving style, the mean value for jerk increased for the overall trip as a result of using pEDAS. Through the feedback obtained from the survey, the participant mentioned difficulty in staying in the optimal speed band and preferred a better way to visualize the deviation from the target speed. In our work, pEDAS indicates the deviation through a varying length of an arrow (Figure~\ref{Fig:a}). If the length is short, drivers were advised not to react abruptly, but rather let the vehicle coast by not engaging the throttle and brake pedals. This feature might not have suited for a few drivers and resulted in the drivers frequently switching between throttle and brake pedals in order to follow the suggestions by pEDAS, which led to an increase in the overall acceleration and deceleration values resulting in higher energy consumption. In our work, the classification of drivers was based on the information provided by them in the survey questionnaire. This could lead to a bias as drivers might choose a style which is quite different from their actual style. For future work, drivers should be classified based on their driving characteristics data obtained from their test runs.
	
	In the survey conducted as part of the subjective analysis, three feedback questions were asked to the participants which addressed the key areas of pEDAS (Table~\ref{appendix:tab3}). The feedback indicated that the participants have found suggestions from pEDAS to be very helpful in not only minimizing the energy consumption in a battery electric vehicle but also additional features such as green-wave, displaying of traffic light remaining time and speed adaptation to a sharp curve were complimentary. The participants were also satisfied with the positioning and size of pEDAS visualization on the heads-up display. However, the feedback also indicated potential areas of improvements. Although the driving route was highlighted in the virtual driving environment, a few drivers missed the turns in the driving route while focusing on the suggestions from pEDAS. Drivers suggested in the future to display a navigation map on the HUD or provide auditory feedback for navigation. Another feedback from participants is that the suggestions from pEDAS to slow down before a stop sign or a sharp turn could appear a bit earlier, so that they had enough time to react. In the driving simulator experiments, the drivers visualized the virtual driving environment on a flat TV monitor with a restricted field of view. Therefore, in the urban environments this resulted in unsafe scenarios at sharp turns. In future to provide an increased field of view to the driver, displaying the virtual environment using either a curved screen, triple-monitor or a virtual reality headset can be considered. 
	
	Furthermore, a survey questionnaire based on TAM and TPB was formulated to understand the factors that influence the user acceptance of the proposed pEDAS. Considering TAM, the perceived usefulness was found to be the strongest predictor of behavioral intention towards using pEDAS. This further implies that persons are likely going to use pEDAS if they think it is useful in their daily commute. The driver’s belief of usefulness of pEDAS is associated with several rewards such as increase in energy savings for the trip, less over speeding, spending less time at traffic light signals and continuous eco-driving assistance throughout the trip. These rewards or performance benefits will in turn contribute to the intention of using pEDAS, thereby increasing the acceptance amongst users. This study found that the perceived ease of use is a significant positive predictor of perceived usefulness. This implies that there is a high probability that the drivers will find pEDAS to be useful if they find it easy to use. This is also logical in a way that if users find it easy to perform a behaviour, here understanding suggestions from pEDAS, it will motivate them to use the technology more often. The analysis also showed that the perceived usefulness is a significant predictor of attitude towards behavior. Attitude, which relates to a person’s positive and negative feelings towards performing a certain behaviour (here, to use pEDAS) is explained by the person’s belief of usefulness. The result implies that if a person believes that pEDAS is useful, then that person will likely develop a positive attitude towards using it. Moreover, the study points out that the attitude, which is a common construct in both TAM and TPB, did not exhibit a significant contribution in predicting the behavior intention. A possible explanation of the aforementioned discrepancies can be inferred from the participant feedback that pEDAS might have created ambiguity and distraction among few drivers. This claim can be further supported from the neutral responses obtained for items 7 and 8 of the Attitude construct and items 2,3 of the PEoU construct as illustrated in Table~\ref{tbl3}. For TPB, the subjective norm was found to have a small explainable variance on the behavioral intention, however, was found to be not statistically significant. This implies that the opinion of important people in one's social network have a very little effect on their intention to use pEDAS. Moreover, it was found from the study that the perceived behavioral control was the strongest predictor of behavioral intention in TPB. This implies that the individuals who perceive ease in using pEDAS in the presence of necessary resources are most likely to intend to use pEDAS in their daily life.  
	\section{Conclusion}
	With the primary objective to improve the drivers' driving styles and thereby reduce the energy consumption in battery electric vehicles, the predictive eco-driving assistance system (pEDAS) proposed in this work provides a continuous feedback to the driver in a visual and auditory manner. The optimal speed suggestions were provided to assist drivers in urban and highway driving environments in the presence of dense stochastic traffic, traffic light signals, traffic signs and curved roads. Our present study investigated the performance of pEDAS and evaluated the acceptance of the proposed system by conducting user studies on a dynamic driving simulator. The objective analysis revealed that pEDAS has the potential to reduce energy consumption in both urban and highway environments for drivers with different driving styles. In addition to that, pEDAS assisted drivers in lowering the speed limit violations and avoiding unnecessary stops at the signalized intersections. Furthermore, the current study used the technology acceptance model (TAM) and theory of planned behavior (TPB) to investigate the influence of the constructs perceived usefulness, perceived ease of use, attitude towards behavior, subjective norms, and perceived behavioral control on the behavioral intention to use pEDAS. The results revealed that the perceived usefulness and perceived behavioral control are the strongest predictors of behavioral intention in using the proposed pEDAS. Additional investigations with a larger sample size must be made in the future to assess the moderating effects due to age, gender, experience and voluntariness while evaluating the user acceptance of pEDAS. In the future work, efforts are targeted on determining and compensating the driver's reaction delay in order to improve the optimal speed tracking ability of the drivers, which can in turn help to achieve additional energy savings. The proposed predictive EDAS can be applied for conventional internal combustion engine (ICE) vehicles as well. The problem formulation, however, must be adapted to include gear shifting and engine ON/OFF strategy as presented in our previous work \citep{Chada2021_2}. Furthermore, the impact on the energy consumptions for the network-wide installations of the proposed pEDAS is a worth while investigating aspect for the future work.
	\section{Acknowledgement}
	The authors would like to extend their gratitude to Yannick Ranker for his support during the driving simulator experiments and all the volunteers at the University of Kaiserslautern. The project is funded by the Ministry for Education and Research (BMBF) of the Federal Republic of Germany and partially supported by the Center of Commercial Vehicle Technology (Zentrum für Nutzfahrzeugtechnologie, ZNT) at the University of Kaiserslautern funded by the Ministry for Science and Health (MWG) of the Federal State of Rhineland-Palatinate.
\\ \\   
	
	\appendix
	\section{Appendix: Survey Questionnaire}
	\label{appendix_A}
	\subsection{Evaluation of user acceptance of a predictive eco-driving assistance system (pEDAS)}
	The survey items under each construct of the technology acceptance model (TAM) and theory of planned behavior (TPB) are listed below. These items are adapted from the work \cite{Voinea2020}.
		\begin{table}[width=1\linewidth,pos=h]
		\caption{Survey Questionnaire - Attitude toward behaviour}\label{appendix:tab2}
		\centering
		\begin{adjustbox}{width=\columnwidth,center}
			\begin{threeparttable}
				\begin{tabular}{>{\arraybackslash}p{1.1\textwidth}}
					I find the pEDAS when I am driving to be:        \\					
					\begin{tabular}{|c|c|c|c|c|c|c|c|}
						\hline
						& & & & & & &\\[-0.8em]
						& & 1 & 2 & 3 & 4 & 5 & \\
						& & & & & & &\\[-0.8em]
						\hline
						& & & & & & &\\[-1em]
						ATT1 & Bad & $\bigcirc$ & $\bigcirc$ & $\bigcirc$ & $\bigcirc$ & $\bigcirc$ & Good\\
						\hline
						& & & & & & &\\[-1em]
						ATT2 & Useless & $\bigcirc$ & $\bigcirc$ & $\bigcirc$ & $\bigcirc$ & $\bigcirc$ & Useful\\
						\hline
						& & & & & & &\\[-1em]
						ATT3 & Undesirable & $\bigcirc$ & $\bigcirc$ & $\bigcirc$ & $\bigcirc$ & $\bigcirc$ & Desirable\\
						\hline
						& & & & & & &\\[-1em]
						ATT4 & Ineffective & $\bigcirc$ & $\bigcirc$ & $\bigcirc$ & $\bigcirc$ & $\bigcirc$ & Effective\\
						\hline
						& & & & & & &\\[-1em]
						ATT5 & Drowsy & $\bigcirc$ & $\bigcirc$ & $\bigcirc$ & $\bigcirc$ & $\bigcirc$ & Alerting\\
						\hline
						& & & & & & &\\[-1em]
						ATT6 & Unpleasant & $\bigcirc$ & $\bigcirc$ & $\bigcirc$ & $\bigcirc$ & $\bigcirc$ & Pleasant\\
						\hline
						& & & & & & &\\[-1em]
						ATT7 & Annoying & $\bigcirc$ & $\bigcirc$ & $\bigcirc$ & $\bigcirc$ & $\bigcirc$ & Not at all annoying\\
						\hline
						& & & & & & &\\[-1em]
						ATT8 & Irritating & $\bigcirc$ & $\bigcirc$ & $\bigcirc$ & $\bigcirc$ & $\bigcirc$ & Likeable\\
						\hline
						& & & & & & &\\[-1em]
						ATT9 & Worthless & $\bigcirc$ & $\bigcirc$ & $\bigcirc$ & $\bigcirc$ & $\bigcirc$ & Assisting\\
						\hline
					\end{tabular} 
				\end{tabular}
			\end{threeparttable}
		\end{adjustbox}
	\end{table}
	
	\begin{table}[width=1\linewidth,pos=h]
		\caption{Survey Questionnaire}\label{appendix:tab1}
		\centering
		\begin{adjustbox}{width=\columnwidth,center}
			\begin{threeparttable}
				\begin{tabular}{>{\centering\arraybackslash}p{0.2\textwidth}>{\arraybackslash}p{0.9\textwidth}}
					\toprule
					Items   & Survey Questionnaire   \\
					\midrule
					&	\begin{tabular}{|c|c|c|c|c|c|c|}
						\hline
						& & & & & &\\[-0.8em]
						& 1 & 2 & 3 & 4 & 5 & \\
						& & & & & &\\[-0.8em]
						\hline
						& & & & & &\\[-1em]
						Strongly disagree & $\bigcirc$ & $\bigcirc$ & $\bigcirc$ & $\bigcirc$ & $\bigcirc$ & Strongly agree\\
						\hline
					\end{tabular} \\
					\textbf{Perceived Usefulness} &            \\
					PU1     &  Using the pEDAS would improve my driving performance      \\
					PU2      &  Using the pEDAS would improve my energy savings      \\
					PU3       &  Using the pEDAS would enable me to react to unsafe driving situations more quickly       \\
					PU4        &  I assume that an pEDAS equipped car would be useful in my daily life        \\
					\textbf{Perceived Ease of Use} &             \\
					PEoU1     & I find it easy to interpret the suggestions given by pEDAS     \\
					PEoU2      & Interacting with the pEDAS would not require a lot of cognitive effort      \\
					PEoU3       &  I find the pEDAS to be distracting while driving       \\
					PEoU4        &  Learning to use the pEDAS would be easy for me        \\
					\textbf{Perceived Behavioural Control}             \\
					PBC1     & I have control over using pEDAS     \\
					PBC2      & I do have the knowledge necessary to use pEDAS      \\
					PBC3       & Given the resources, opportunities and knowledge it takes to use pEDAS, it would be easy for me to use it in my daily commute       \\ 
					\textbf{Subjective Norms}             \\
					SN1     & People whose opinions I value would think that I should use pEDAS     \\
					SN2      & People who influence my behaviour would think that I should pEDAS      \\
					\textbf{Behavioural intention}             \\
					BI1     & If the pEDAS is available in the market at an affordable price, I intend to purchase this feature    \\
					BI2      & Assuming that the pEDAS is available in my car, I intend to use it regularly when I am driving \\
					\bottomrule
				\end{tabular}
			\end{threeparttable}
		\end{adjustbox}
	\end{table}
	
	\begin{table}[width=1\linewidth,pos=h]
		\caption{Feedback questionnaire}\label{appendix:tab3}
		\centering
		\begin{adjustbox}{width=\columnwidth,center}
			\begin{threeparttable}
				\begin{tabular}{>{\centering\arraybackslash}p{0.2\textwidth}>{\arraybackslash}p{0.9\textwidth}}
					\toprule
					Items   & Feedback questions   \\
					\midrule
					Q1     & Please give your feedback on whether the positioning/size of the pEDAS visualization on the heads-up display is appropriate    \\
					Q2      & Are you able to guess the reasons behind the suggestions given by pEDAS? Were there any situations where the suggestions given by pEDAS were ambiguous?     \\
					Q3      &  Please give your feedback if you would like to have some additional suggestions to be given by pEDAS     \\
					\bottomrule
				\end{tabular}
			\end{threeparttable}
		\end{adjustbox}
	\end{table}

	\bibliographystyle{cas-model2-names}
	
	\bibliography{cas-refs}
	
	%
	%
	%
	%
	
\end{document}